\documentstyle[12pt,aaspp4,psfig]{article}
\begin{document}

\title{Two-Phased Intra-Cluster Medium in the 
Centaurus Cluster of Galaxies} 

\author{Yasushi Ikebe}
\affil{Max-Planck-Institut f\"{u}r extraterrestrische physik,
Postfach 1603, D-85740, Garching, Germany}

\author{Kazuo Makishima\altaffilmark{1} and Yasushi Fukazawa}
\affil{Department of Physics, University of Tokyo, 
7-3-1 Hongo, Bunkyo-ku, Tokyo 113-0033, Japan}

\author{Takayuki Tamura}
\affil{Institute of Space and Astronautical Science,
3-1-1 Yoshinodai, Sagamihara, Kanagawa 229-8510, Japan}

\author{Haiguang Xu}
\affil{Institute for Space and Astrophysics, 
Department of Applied Physics, School of Science,
Shanghai Jiao Tong University, 1954 Huashan Road,
Shanghai, 200030, PRC}

\and

\author{Takaya Ohashi and Kyoko Matsushita}
\affil{Department of Physics, Tokyo Metropolitan University,
1-1 Minami-Ohsawa, Hachioji, Tokyo 192-0397, Japan}

\altaffiltext{1}{also Research Center for the Early Universe (RESCUE),
University of Tokyo, 7-3-1 Hongo, Bunkyo-ku, Tokyo 113-0033, Japan}

\newpage
\begin{abstract}
{\it ASCA} and {\it ROSAT} data of the Centaurus cluster,
holding a cD galaxy NGC~4696, were analyzed, 
as a partial re-analysis after Fukazawa et al. (1994),
Fabian et al. (1994), and Allen \& Fabian (1994).
Radial brightness profiles in different energy bands show
that the central excess emission of this cluster,
known previously in soft X-rays ($<4$ keV),
is present also in the hard energy band up to 10~keV.
Therefore the central excess emission cannot be explained 
solely by a temperature drop together with
a King-type potential having a flat core,
and requires a deeper central potential.
A double-$\beta$ brightness distribution gives a good account
of the data.
A deprojected energy spectrum within a spherical region
of radius $\sim 30$ kpc at the cluster center indicates
that the ICM cannot be isothermal there.
Simultaneous fitting of the three (PSPC, GIS, and SIS) energy spectra 
extracted from the central region within a projected radius of 3 arcmin
showed that a two-temperature model that has temperatures of 1.4~keV 
and 4.4~keV and a common metallicity of 1.0 solar with the Galactic 
absorption gives the best description to the spectra.
A cooling-flow spectral model did not give satisfactory fit.
These results reconfirm the two-phase picture developed
by Fukazawa et al. (1994) in which
the hot ICM fills nearly the entire cluster volume,
wherein a small amount of cooler plasma is intermixed 
and localized near the cluster center.
A 3-dimensional cluster model incorporating the two-phase picture,
the double-$\beta$ brightness distribution,
and the central metallicity increase
reproduced the overall {\it ASCA} and {\it ROSAT} data successfully.
The spatial distribution of the dark matter that is derived by
subtracting the stellar mass from the calculated total-gravitating mass
deviates from a King-type profile and exhibits a central excess.
Another two-phase cluster model in which the dark-matter density
profile is given with
the {\it universal density profile} (Navvaro, Frenk, \& White 1996,1997)
also gave a satisfactory account to the data.
A detailed comparison of the iron mass distribution
with that of the stellar component reveals that
the iron is more widely spread than stars, which may have been caused
by energetic supernovae and the motion of the cD galaxy.
Since the derived characteristics of the cool phase including 
the temperature, angular extent, gas mass, and metallicity,
are on a smooth extension of those of inter-stellar medium (ISM)
of elliptical galaxies,
the cool phase can be regarded as the ISM associated with the cD galaxy,
while the high X-ray luminosity of the cool phase 
($1.1\times10^{43}$ ergs s$^{-1}$ in 0.5--4.0 keV)
is interpreted as a result of a compression by the surrounding
hot phase.
The cool-phase X-ray emission is presumably sustained by
energies dissipated by infalling mass to the cD galaxy
in the course of cluster evolution.
\end{abstract}


\newpage
\section{Introduction}
A central dominant galaxy (cD galaxy) is frequently 
found at the center of a galaxy cluster.
Previous imaging X-ray observations revealed
that X-ray emission from intra-cluster medium 
(ICM; e.g. Sarazin 1988) often exhibits strongly peaked brightness,
which coincide with cD galaxies (e.g. Jones \& Forman 1984).
Such clusters can be classified as XD clusters (Jones \& Forman 1984) 
or cD clusters.
This phenomenon has generally been interpreted in terms of 
a temperature gradient in the ICM, particularly due to cooling flows 
(CF; see Fabian 1994 for a review) occurring at the densest part of 
the ICM where the radiative cooling time is considerably shorter than 
the Hubble time.
Due to limited spectroscopic capabilities of pre-{\it ROSAT} instruments, 
however, the radial temperature profiles of the ICM were usually 
estimated indirectly from the observed X-ray brightness profiles,
assuming a hydrostatic equilibrium of the ICM within a plausible shape 
of the gravitational potential
(i.e. deprojection method; Fabian et al. 1981).
Results of such calculations indicate that the ICM temperature often 
decreases towards the center, typically down to $\sim 1$ keV.
The existence of such a cool component was spectroscopically confirmed 
with the {\it Einstein} SSS and FPCS (Canizares et al. 1979; 
Mushotzky et al. 1981; Lea et al. 1982; White et al. 1994).
In addition, excess absorption was observed with the {\it Einstein} 
SSS from central regions of a number of cD clusters
(White et al. 1991; Johnstone et al. 1992).
With the advent of {\it ROSAT} (Tr\"{u}mper 1984),
the temperature gradient decreasing towards the cluster center
have been detected in many cD clusters (e.g. Allen et al. 1993).

In reality, the ICM temperature structure is thought to be quite 
multi-phased, in such a way that there exist many temperatures and 
densities at a given radius 
(e.g. inhomogeneous cooling flows; Nulsen 1986).
Actually, such a multi-phase structure was suggested 
in the central region of the Virgo cluster, through fine-resolution 
spectroscopy with the {\it Einstein} FPCS (Canizares et al. 1982).
Nevertheless, single-phase approximations, 
in which a single temperature is assigned to each radius, 
have been used extensively as an approximation.
This description has successfully explained the observed data,
at least in energy bands below 3~keV, 
yielding results consistent with those from a multi-phase modeling
(Thomas, Fabian, \& Nulsen 1987).
However, the single-phase approximation 
may no longer be valid in higher energy ranges,
and may fail to reveal some important aspects of cD clusters.

Another key issue related to the central regions of clusters 
is spatial distribution of the ICM metallicity, 
which directly reflects the evolutionary history 
of the cluster and its member galaxies. 
Koyama, Takano, \& Tawara (1991), through scanning observations 
of the Virgo cluster with the {\it Ginga observatory}, 
showed that the iron abundance of the ICM 
clearly increases towards its cD galaxy, M87.
White et al. (1994), utilizing the different sizes of
fields of view of the {\it Einstein} SSS and the {\it Ginga} LAC,
revealed the central metal concentration in A496, A2142,
and possibly in A1795 and A2199.
However, metallicity gradients in other clusters remained unknown
or inconclusive (e.g. for the Perseus cluster; Mushotzky et al. 1981;
Ponman et al. 1990; Kowalski et al. 1993; Arnaud et al. 1994).
This issue was rather hard to explore even with {\it ROSAT},  
because of the limited spectral resolution 
and the limited energy band (0.1--2.4 keV) 
that does not cover the Fe-K line energy (6--7 keV).

The {\it ASCA observatory} (Tanaka, Inoue, \& Holt 1994), 
with its superior imaging spectroscopic capability up to 10 keV, 
has drastically improved our knowledge on the issues of 
temperature structure and metallicity distribution in ICM.
Among many clusters so far observed with {\it ASCA},
the Centaurus cluster (Cen cluster for short) is of particular interest:
it exhibits both the metallicity concentration 
and the central cool emission in a most outstanding manner,
as already reported by Ohashi et al. (1994), 
Fukazawa et al. (1994; hereafter FEA94), 
and Fabian et al. (1994; hereafter FABT94).
The phenomena are so fascinating 
that they deserve a still more detailed investigation. 

We have hence re-analyzed the previously and the newly 
obtained {\it ASCA} data of the Cen cluster,
together with those from the {\it ROSAT} PSPC 
having a better angular resolution.
For the {\it ASCA} data analysis,
we have taken fully into account the complex response 
of the X-Ray Telescope
(XRT; Serlemitsos et al. 1995; Tsusaka et al. 1995; Takahashi et al. 1995)
on board {\it ASCA} 
using the method described in Ikebe et al. (1997).
We have reconfirmed the conclusion made by FEA94,
that the ICM in the central region of the Cen cluster 
is made up of hot and cool phases,
and investigated their detailed characteristics.
In section 2, we briefly review past X-ray results on the Cen cluster,
and point out relevant scientific issues.
Section 3 gives a brief description of the {\it ASCA} and {\it ROSAT}
observations.
Results of the new analysis are presented in sections 4 through 6,
followed by discussion in section 7.
We assume  $H_{\rm 0} = 50$ km s$^{-1}$ Mpc$^{-1}$ throughout.

\section{Previous X-ray Results and Scientific Issues}
The Cen cluster, or A3526, is a relatively poor cluster of galaxies 
(e.g. Dickens, Currie, \& Lucey 1986) at a redshift of $z=0.0104$.
It is one of the nearest and the most well studied Abell clusters.  
At the cluster center, there exists a cD galaxy NGC~4696. 
The Cen cluster has been observed repeatedly in X-rays, 
with {\it Uhuru} (Giacconi et al. 1972; Forman et al. 1978),
{\it HEAO-1} (Mitchell \& Mushotzky 1980), 
the {\it Einstein Observatory} 
(Matilsky, Jones, \& Forman 1985; David et al. 1993), 
{\it EXOSAT} (Edge \& Stewart 1991), 
and {\it Ginga} (Yamashita 1992).
These observations indicate the 2--10 keV luminosity 
of $7 \times 10^{43}$ erg s$^{-1}$, 
and the ICM temperature of $\sim 4 $ keV except in the central region. 
A central excess brightness, interpreted as a CF,
was detected within $6'$ of the center
(Matilsky et al. 1985; Edge \& Stewart 1991).

Allen \& Fabian (1994; hereafter AF94) analyzed the 
radius-sorted {\it ROSAT} PSPC spectra of the Cen cluster,
and found that the projected ICM temperature
significantly decreases within $\sim 6'$ of the cluster center, 
where they also found some evidence for 
increased metallicity or excess absorption.
A single-phase deprojection solution derived by AF94
implies that the ICM temperature starts decreasing from $\sim 3.8$ keV 
at a radius of $\sim 6'$, roughly as $\propto R^{0.5}$
as a function of the three-dimensional radius $R$,
reaching $\sim 1$ keV at the cluster center.
They derived a mass deposition rate of 30--40 $M_{\odot}$ yr$^{-1}$.

Subsequently, the Cen cluster was observed with {\it ASCA}.
The {\it ASCA} spectra within a central $\sim8'$ region 
cannot be fit with a single temperature model, 
requiring at least two temperatures of $\sim 4$~keV 
and $\sim 1$~keV (Ohashi et al. 1994; FEA94; Ikebe 1995).
FABT94 analyzed the spectrum from the central region 
of the Cen cluster obtained in the same observation,  
and spectroscopically estimated the mass deposition rate to be
15--20 $M_{\odot}$ yr$^{-1}$ in terms of homogeneous-CF approximation.   
Thus, the available X-ray data consistently indicate presence 
of the cool region at the center of the Cen cluster.
The {\it ASCA} data also revealed a drastic metallicity increase 
towards the center (Ohashi et al 1994; FEA94; Ikebe 1995).
This occurs within $\sim8'$ of the cD galaxy NGC~4696
and spatially coincides with the appearance of the cooler component.
The {\it ASCA} data of the Virgo cluster also show qualitatively similar
temperature and metallicity profiles (Matsumoto et al. 1996).

Based on these new results, FEA94, Makishima (1994), and Ikebe (1995) 
developed a two-phased modeling of the ICM,
wherein a cool component is confined within $\sim 5'$ of the center,
immersed in a hot component that fills the entire cluster volume.
This view is apparently supported by the fact 
that the two temperatures, $\sim 4$~keV and $ \sim 1$~keV,
are quite constant as a function 
of the projected radius from the center (FEA94; Ikebe 1995).
However, an X-ray spectrum involving many temperature components,
such as is implied by a single-phased ICM after projection, 
or by inhomogeneous CFs,
may well be approximated with just two model temperatures (FABT94).
In addition, the apparent existence of the hot emission at the cluster 
center might be an artifact caused by the complex responses 
of the XRT on board {\it ASCA}.
In short, the true three-dimensional temperature structure of the 
ICM in the central region of this cluster is still unknown.

\section{Observation and Data Selection}
\subsection{{\it ASCA} observations}
\begin{figure}
\centerline{
\psfig{file=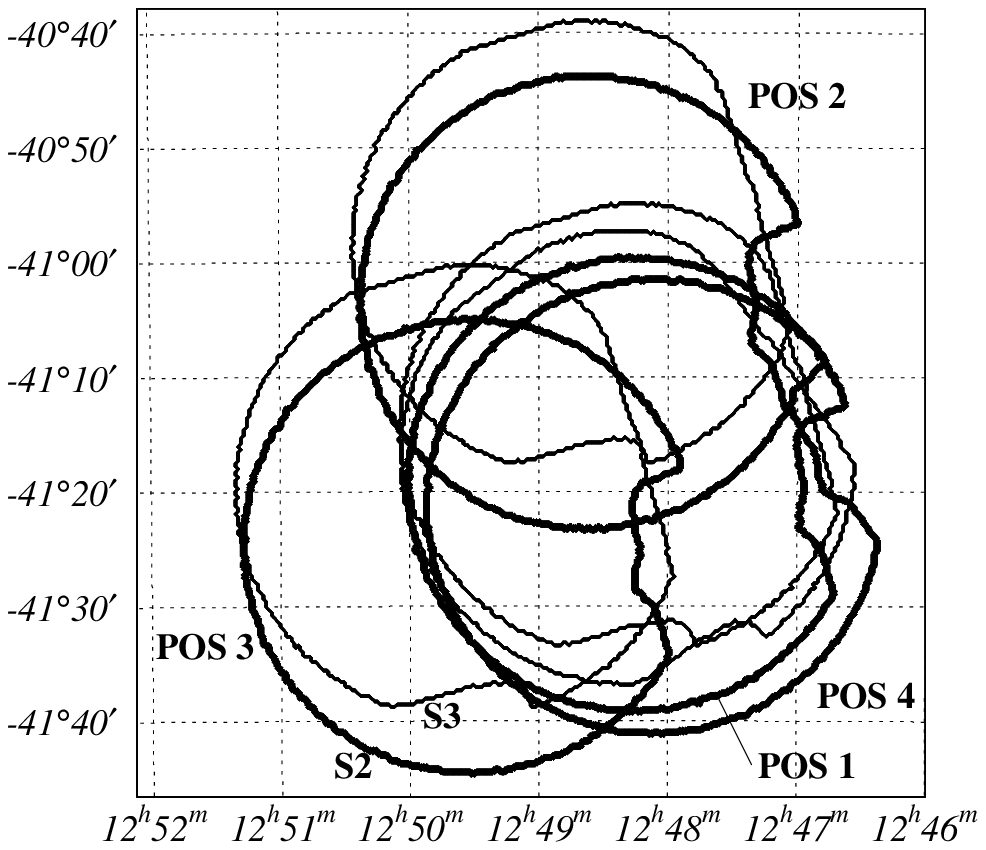,width=9.0cm,angle=0,clip=}
\psfig{file=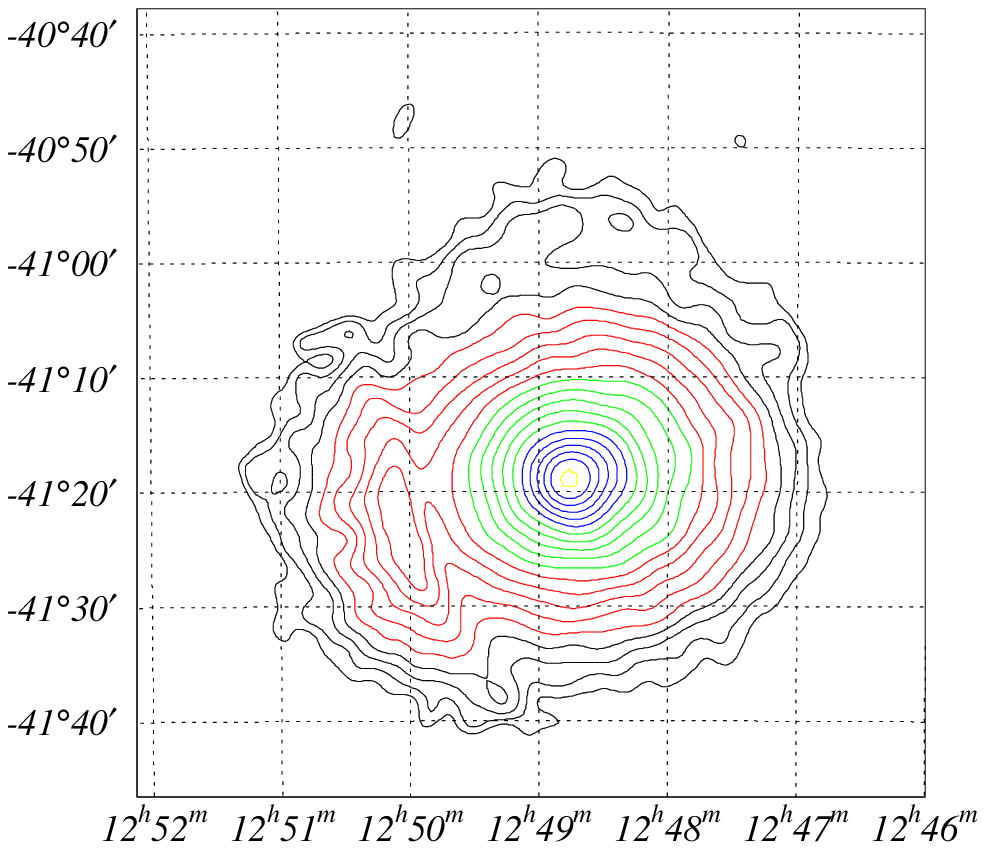,width=9.0cm,angle=0,clip=}}
\figcaption{(a) Field of view of GIS2 and GIS3
for the four pointing positions used in the analysis.
(b) The 0.7-10~keV X-ray intensity contour of the Centaurus cluster 
derived from the GIS data.
The background is subtracted.
A structure seen at ($12^h 50^m, -41^{\circ} 23'$) is an artifact
due to synthesizing the data from the four pointings with 
the exposure correction but without the vignetting correction.
\label{fig:images}}
\end{figure}

The Cen cluster was observed twice with {\it ASCA}, in 1993 and 1995.
The observation log is given in Table~\ref{tab:obslog},
and the pointing positions are shown in Figure~\ref{fig:images}a.

The 1993 observation took place during the PV (performance verification) 
phase.
Because the X-ray emission from the Cen cluster extends beyond 
edge of the $25'$-radius field of view of the {\it ASCA} GIS 
(Gas Imaging Spectrometer; Ohashi et al. 1996; Makishima et al. 1996a),
three separate pointings were made;
one right on the cluster center, 
and the others with $18'$ offset from the cluster center.
These data were already used by FEA94, FABT94 and Ikebe (1995).
The 1995 observation was performed in the AO-3 period,
by pointing on the cluster center for about 59 ksec.

In these observations, the two GIS detectors, GIS2 and GIS3, 
were operated in the normal PH mode (Ohashi et al. 1996).
The two detectors of the SIS (Solid State Imaging Spectrometer) instrument,
SIS0 and SIS1, were operated in 4CCD BRIGHT mode during the 1993 observation,
while in 1CCD FAINT mode during the 1995 observation.
In the following data analysis, we use the GIS data (\S4, \S5 and \S6),
as well as a limited portion of the SIS data (\S5).

\subsection{Selection of the {\it ASCA} data}
We discarded the GIS and SIS data that were taken 
when the XRT view direction was within $5^\circ.0$ of the local horizon.
This is to avoid the periods when the target is earth occulted, 
and to exclude the effect of solar X-rays scattered off the atmosphere.
Additional screening that the elevation angle from the 
sunlit earth is greater than $25^\circ$ and $20^\circ$ 
was applied to the SIS0 and SIS1 data, respectively.
In order to ensure a low and stable particle background, 
we also discarded the GIS and SIS data 
acquired under geomagnetic cutoff rigidity smaller than 6 GeV c$^{-1}$.

The background is composed of cosmic X-ray background (CXB) 
and non X-ray background (NXB).
The CXB component was derived from the data of 
blank-sky observations performed in 1993 and 1994.
We estimated the NXB component by using the data 
taken when the XRT is pointing at the dark (night) earth.
Detailed description of the {\it ASCA} background properties,
and the method of background subtraction, are found in  Kubo et al. (1994),
Ikebe et al. (1995), Ueda et al. (1996), and Ishisaki et al. (1997).

By analyzing the {\it ASCA} data taken in 1995,
we found that the X-ray peak position is systematically 
displaced by $\sim0.6'$ from that measured in 1993.
This is most probably due to inaccuracy in 
the satellite attitude determination (see Gotthelf 1996).
In order to eliminate the discrepancy, 
we have adjusted the individual photon positions of the 1995 data
so that the peak positions obtained in the two epochs become coincident.
This does not affect our results,
since we do not discuss absolute alignment X-ray features
relative to the optical position,
and we are mostly concerned with angular scales larger than $1'$.

Figure~\ref{fig:images}b shows the 0.7--10 keV 
GIS image of the Cen cluster obtained in this way.
Data from the two GIS detectors are combined,
and events from the four partially-overlapping pointings 
(obtained in 1993 and 1995)
are synthesized into a single image after correction for exposure 
and for the systematic attitude difference mentioned above.
The background was subtracted 
but the image is not corrected for the XRT vignetting.
The emission from the cluster is thus detectable
up to the boundary of the GIS field of view,
exhibiting an approximate circular symmetry around the cD galaxy,
which coincides with the X-ray peak.

Although the GIS has sensitivity in the 0.7--10~keV range 
(Ohashi et al. 1996),
its response below $\sim1$~keV is still subject to some uncertainties
arising from complex xenon M-edge structures.
This sometimes causes a noticeable discrepancy 
between the GIS and SIS spectra around 1~keV.
Furthermore the actual GIS image of Cygnus X-1,
which is needed in the image analysis with the GIS data,
is blurred in soft energies by interstellar grain scattering.
As a consequence, the GIS image analysis becomes less reliable 
below $\sim 1.5$ keV.
Throughout this paper, we hence discard the GIS data below 1.6~keV.

\subsection{{\it ROSAT} observations}
In order to complement the limited angular resolution of {\it ASCA},
we also utilize the {\it ROSAT} All Sky Survey data 
and those from pointing observations with the {\it ROSAT} PSPC. 
The {\it ROSAT} All Sky Survey  observations over the Cen cluster
were performed in January 1991 with an average exposure time of 290 sec.
The X-ray emission from the Cen cluster was detected
up to at least $90'$ ($\sim$1.6~Mpc) from the center.
Several {\it ROSAT} pointings onto the Cen cluster
were also performed later with the PSPC. 

Our analysis of the {\it ROSAT} surface brightness 
distribution utilizes the All Sky Survey data,
because the X-ray size of the Cen cluster significantly exceeds 
the PSPC field of view.
The All Sky Survey data provide a virtually flat exposure over the whole 
cluster,
and the point-spread function is averaged out to become nearly position 
independent.
These properties of the All Sky Survey data make the analysis relatively 
simple.
In contrast, for the spectral analysis (\S~5) that requires good photon 
statistics, we use the 7.5 ksec PSPC pointing data
taken on 1992 February 2 right on the cluster center.


\section{Analysis of the Radial Brightness Profiles}

Previous X-ray imaging observatories were
sensitive only to soft X-rays below $\sim$4~keV.
It is therefore of great importance to investigate X-ray brightness 
distributions in broader energy bands using the {\it ASCA} data.
Accordingly, we here analyze the radial count-rate 
profiles obtained with the {\it ASCA} GIS, 
in comparison with that from the {\it ROSAT} PSPC.
For the reason described later in \S 4.2,
we do not analyze the SIS data in this section.

\subsection{Radial count-rate profiles}

\begin{figure}
\psfig{file=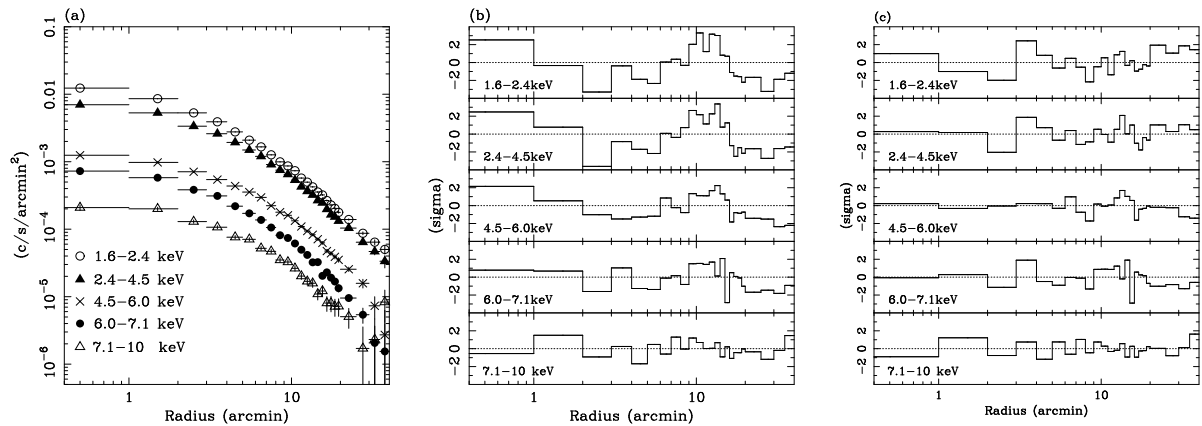,width=16.0cm,angle=-90,clip=}
\figcaption{(a)The {\it ASCA} GIS radial count-rate profiles
in five energy bands;
1.6--2.4~keV, 2.4--4.5~keV, 4.5--6.0~keV, 6.0--7.1~keV, and
7.1--10~keV. The background is subtracted.
(b) The residuals of the single-$\beta$ model fits 
to the GIS radial count-rate profiles.
(c) The residuals of the double-$\beta$ model fits
to the radial count-rate profiles.
The fit parameters refer to Table~\ref{tab:rprofit}.
\label{fig:gisrpro}}
\end{figure}

By azimuthally averaging the data around the X-ray centroid,
we have derived the GIS radial X-ray count-rate profiles.
The data from the two GIS detectors (GIS2 and GIS3)
obtained in the 4 pointings were all summed up.
Figure~\ref{fig:gisrpro}a shows the background-subtracted
GIS profiles thus derived in 5 different energy bands, 
1.6--2.4, 2.4--4.5, 4.5--6.0, 6.0--7.1, and 7.1--10.0~keV.
The lowest energy band is sensitive to the cool component,
while the fourth band covers the Fe-K energy region.

To assist the GIS data analysis,
we also derived a radial count-rate profile 
of the {\it ROSAT} All Sky Survey data, from which 
we eliminated contaminating point sources.
We used the PI channels from 52 to 201, 
roughly corresponding to an energy range of 0.5--2.0~keV.
Figure~\ref{fig:pspcrpro} shows the derived PSPC count-rate profile,
which includes the background.
Thus, we have altogether six radial profiles,
one from {\it ROSAT} and five from {\it ASCA}.


\subsection{Method of analysis}
In order to quantify the derived radial count-rate profiles,
we fit them in a ``forward'' manner.
That is, we represent the three-dimensional emissivity distribution
of the ICM by a spherically-symmetric numerical model
involving a small number of adjustable parameters.
The model emissivity is convolved 
with the image-response matrix (IRM: Ikebe et al. 1997) 
that represents the vignetting effects and 
the position- and energy-dependent 
point-spread function (PSF) of the telescope plus detector system,
and then with the energy response matrices of the focal plane instruments.
The obtained model prediction, to which the expected background is added,
is directly fitted to the observed count-rate profiles.
We search for the best-fit model, through minimization 
of the $\chi^2$ function defined in Ikebe et al. (1997).
This approach allows us to fully take into account 
the known instrumental responses,
and provides a particularly powerful analysis method of the {\it ASCA} data
that are subject to the very complex PSF.
However, we must carefully choose the initial model 
in order to obtain meaningful results.

The PSF of the {\it ASCA} XRT+GIS can be constructed 
using the actual images of Cygnus X-1
(Takahashi et al. 1995; Ikebe 1995; Markevitch et al. 1996).
When fitting the {\it ASCA} GIS profiles,
we apply 2\% systematic errors to the model prediction,
and 10\% systematic errors to the background intensity.
However, Cygnus X-1 is too bright to be observed with the SIS.
Consequently the present analysis method is not applicable to 
the SIS data,
and hence we do not analyze the SIS data in this section.
As to the PSF for the {\it ROSAT} All Sky Survey (XRT+PSPC) data,
we use the one calculated at 1.0~keV and averaged over the whole PSPC
field of view (Hasinger et al. 1994).
The energy dependency of the PSF is small, and negligible
in the following analysis.

The IRM used here has a size of 102 in the sky-region dimension,
both for the GIS and the PSPC.
The 102 regions comprise 16 shell regions of $0'.25$ width
covering a spherical region within $R=4'$,
where $R$ is three-dimensional radius from the center,
and 86 shell regions of $1'$ width covering $R=4'-90'$.
Thus, we implicitly assume that no X-ray emission 
comes from outside the maximum radius of $R_{\rm max}= 90'$
as is justified by the {\it ROSAT} PSPC image.

\subsection{Single-$\beta$ fits}

As the three-dimensional emissivity model,
we first try a single-$\beta$ profile defined as
\begin{equation} \label{eq:1beta}
  \epsilon(R) = 
  \epsilon_0 \left[1+ \left(\frac{R}{R_{\rm c}}\right)^2
\right]^{-3\beta}~~,
\end{equation}
where $R$ is again the three-dimensional radius,
$R_{\rm c}$ is the core radius, and $\beta$ is the beta-parameter.
The energy spectrum is assumed to be that of thermal X-ray 
emission from an isothermal plasma of temperature 4.0 keV 
with a constant metallicity of 0.3 solar,
modified by Galactic absorption of 
$N_{\rm H}$ = 8.8 $\times 10^{20}$ cm$^{-2}$ (Stark et al. 1992).
Although this assumption is inconsistent with the 
significant spectral changes seen from the Cen cluster, 
its effect is negligible in the present analysis
because we here {\it individually} treat count-rate profiles 
integrated over relatively narrow energy bands.

\begin{figure}
\psfig{file=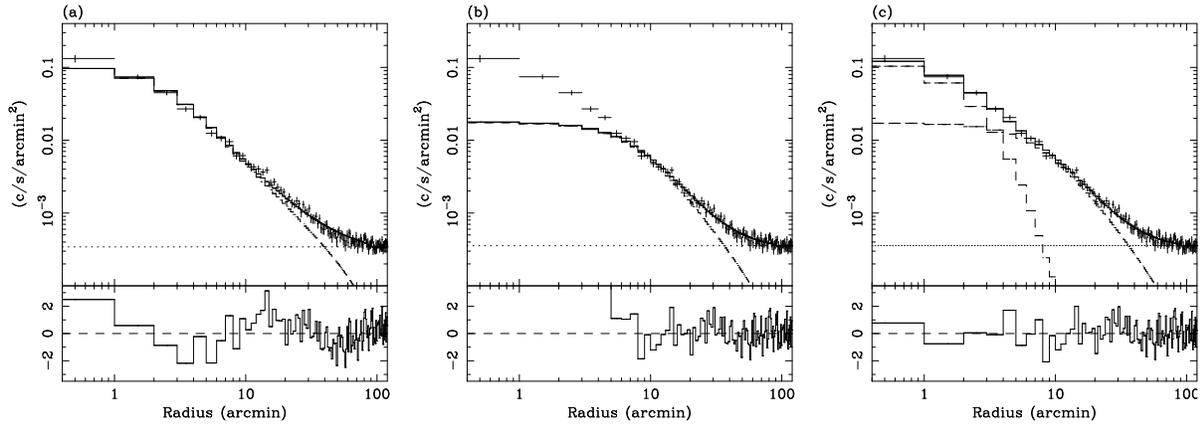,width=16.0cm,angle=-90,clip=}
\figcaption{The {\it ROSAT} PSPC radial count-rate profile
in the 0.5-2.0~keV band. The background is included.
(a) The single $\beta$-model fits to the PSPC radial count-rate
profile of the whole $0-120'$ region and (b) of the $6'-120'$ region.
(c) The double $\beta$-model fit to the PSPC radial count-rate 
profile.
The fit parameters refer to Table~\ref{tab:rprofit}.
In the fits, a constant but free background is added.
\label{fig:pspcrpro}}
\end{figure}

We thus fitted the PSPC profile over a range of $r=0'-120'$
with the single-$\beta$ model plus a constant background,
where $r$ denotes the projected radius.
As presented in Table~\ref{tab:rprofit} and Figure~\ref{fig:pspcrpro}a,
the fit was unacceptable,
because of the long-known central excess emission 
(Matilsky et al. 1985; AF94).
By limiting the fit range to $r>6'$
where the ICM is virtually isothermal (AF94; FEA94),
the fit has become acceptable 
(Table~\ref{tab:rprofit}, Fig.~\ref{fig:pspcrpro}b)
with the best-fit model revealing the strong central excess above it.
The obtained best-fit parameters (Table~\ref{tab:rprofit}) 
are close to those ($R_{\rm c}=5'\pm1'$ and $\beta=0.45\pm0.03$)
derived previously by Matilsky et al. (1985), 
who used the 0.5--3.5~keV {\it Einstein} IPC data 
over a narrower range of  $r=6'-20'$.

Similarly, we fitted the GIS radial count-rate profiles 
with the single-$\beta$ model over $r=0'-40'$.
The results are shown in Table~\ref{tab:rprofit} and 
Figure~\ref{fig:gisrpro}b.
Thus, the single-$\beta$ model failed to reproduce 
the GIS profiles in the lower four energy bands.
The failure is thought to be caused by the central excess emission,
because the fit residuals behave qualitatively in the same manner
as those in the unacceptable single-$\beta$ fit to the PSPC profile 
using the entire radius range.
We hence conclude that the strong central-excess emission of the Cen
cluster,
previously known in X-ray energies below $\sim 4$ keV, 
is present in harder energy ranges at least up to 7.1 keV,
although the excess may be rather energy dependent.

The GIS radial profile in the highest energy band (7.1--10~keV)
was fitted apparently well with a single-$\beta$ model.
However, the obtained $\beta$ model is unusually flat with $\beta=0.44$,
and is inconsistent with that derived with the PSPC profile outside
$6'$ ($\beta=0.57$; Table~\ref{tab:rprofit}).
If the 7.1--10~keV brightness profile in the outer cluster region
were really flatter than that in 0.5--2~keV, 
the ICM temperature would rise significantly towards outside,
e.g., from 3~keV at $R\sim10'$ to 5~keV at $R\sim60'$,
and 7~keV at $R\sim90'$;
this disagrees with the previous observational results (AF94; FEA94).
Such an unusually flat solution often occurs
when we force a single-$\beta$ model to fit an X-ray 
radial profile having a strong central excess (Makishima 1995).
We hence infer that the central excess emissivity is
present even in the highest GIS energy band.
This inference is supported by the fact
that the single-$\beta$ solution in 7.1--10 keV
has nearly the same shape as those unacceptable models 
which approximate the 2.4--4.5, 4.5--6.0, and 6.0--7.1 keV profiles,
as can be seen in Table~\ref{tab:rprofit}.
If the data statistics were somewhat better,
the single-$\beta$ fit to the 7.1--10 keV GIS profile 
would have been rejected as well.
A further confirmation of these results is presented below.


\subsection{Double-$\beta$ fits}
To better quantify the count-rate profiles,
we proceed to use a double $\beta$-model given as
\begin{equation} \label{eq:2beta}
 \epsilon(R) 
 = \epsilon_{0,1}
    \left[1+\left(\frac{R}{R_{\rm c,1}} \right)^2 \right]^{-3\beta_1}
  +\epsilon_{0,2}
    \left[1+\left(\frac{R}{R_{\rm c,2}} \right)^2 \right]^{-3\beta_2}~~.
\end{equation}
Here $\epsilon_{0,1}$, $\epsilon_{0,2}$, $R_{\rm c,1}$, $R_{\rm c,2}$
$\beta_1$, and $\beta_2$ are model parameters;
the term with suffix 1 (wider $\beta$-component)
represents the overall cluster emission,
while that with suffix 2 (narrower $\beta$-component)
represents the central excess emission.
This model has been applied successfully to the {\it ASCA} GIS data of 
the Fornax cluster (Ikebe et al. 1996), 
the Hydra-A cluster (Ikebe et al. 1997),
Abell 1795 (Xu et al. 1998), and NGC~4636 (Matsushita et al. 1998).
Furthermore, Mulchaey \& Zabludoff (1998) used the same model
to quantify the {\it ROSAT} PSPC count-rate profiles of several galaxy
groups.

Like in the previous subsection, the model is assumed to have
the energy spectrum for the isothermal plasma of 4.0~keV temperature 
and 0.3~solar metal abundance.
The core radius and beta value of the wider $\beta$-component,
are fixed at $R_{\rm c,1}$=$7'.3$ and $\beta_1$=0.57, respectively,
as obtained with the $6'-120'$ PSPC profile,
while the other 4 parameters are left free.
This procedure is justifiable,
because the observed near-isothermality of the ICM emission for $R>6'$
requires the outer-region emissivity profile 
to be independent of X-ray energy,
although the amplitude of the central excess may be rather energy dependent.
Furthermore in our final analysis made in \S~6, 
we let $R_{\rm c,1}$ and $\beta_1$ to vary freely.

\begin{figure}
\psfig{file=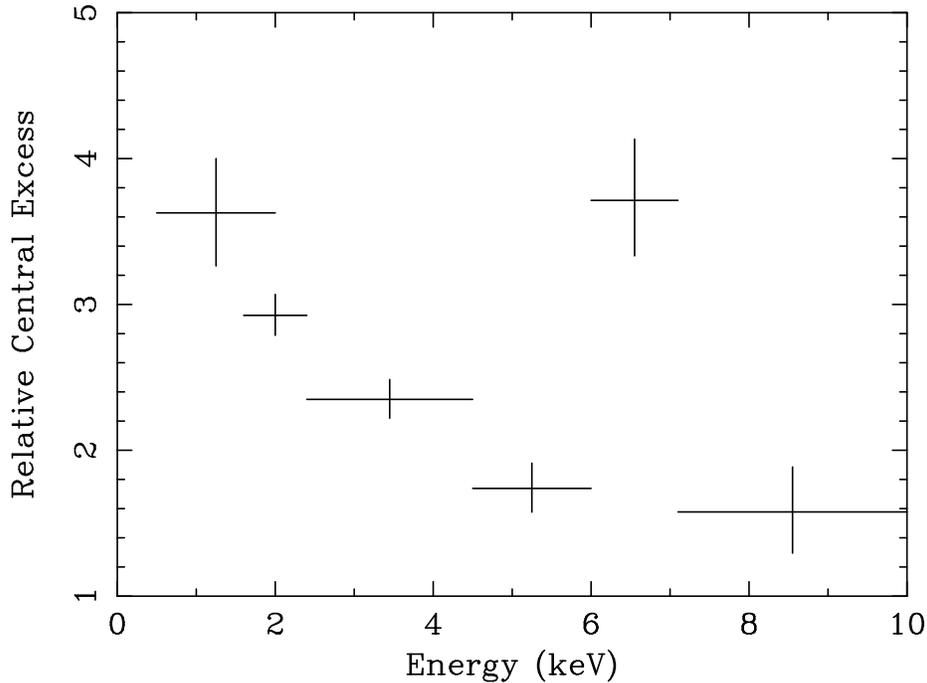,width=15.0cm,angle=-90,clip=}
\figcaption{Relative central excess (see text) in the six energy
bands.
The vertical bars represent 90\% ($\Delta \chi^2$=2.706) errors.
\label{fig:rce}}
\end{figure}

As summarized in Table~\ref{tab:rprofit} and 
illustrated in Figure~\ref{fig:pspcrpro}c and \ref{fig:gisrpro}c,
this double-$\beta$ model has given acceptable fits to 
most of the GIS and PSPC count-rate profiles,
possibly except in the 1.6--2.4~keV band.
The obtained values of $R_{\rm c,2}$ and $\beta_2$ are
roughly energy independent within rather large errors.
In contrast, as shown in Figure~\ref{fig:rce},
we observe a significant energy dependence 
in relative strength of the central excess emission, 
or ``relative central excess'';
this has been calculated in each GIS or PSPC energy band 
by integrating the flux implied by the narrower $\beta$-component
and normalizing it to that carried by the wider $\beta$-component, 
both calculated within a sphere of $R = 5'$.

 From these results, we conclude
that the central excess emission of the Cen cluster 
results from a combination of three apparently separate effects.
The first is the presence of the spectral cool component (FEA94; FABT94),
which causes the prominent increase in the relative central excess 
below $\sim 4$ keV.
The second effect is the metallicity increase towards the center,
which makes the relative central excess significantly higher
in the Fe-K energy band than in the adjacent energy ranges.
The somewhat worse fit in the 6.0--7.1 keV band is
probably due to an additional effect from the metallicity gradient.
The last effect is the intrinsic shape of the gravitational potential,
which is inferred since the central excess 
does not vanish in Figure~\ref{fig:rce} even in the highest energy band,
where the former two effects are expected to be negligible.


\subsection{Deprojected spectrum at the center}

\begin{figure}
\centerline{\psfig{file=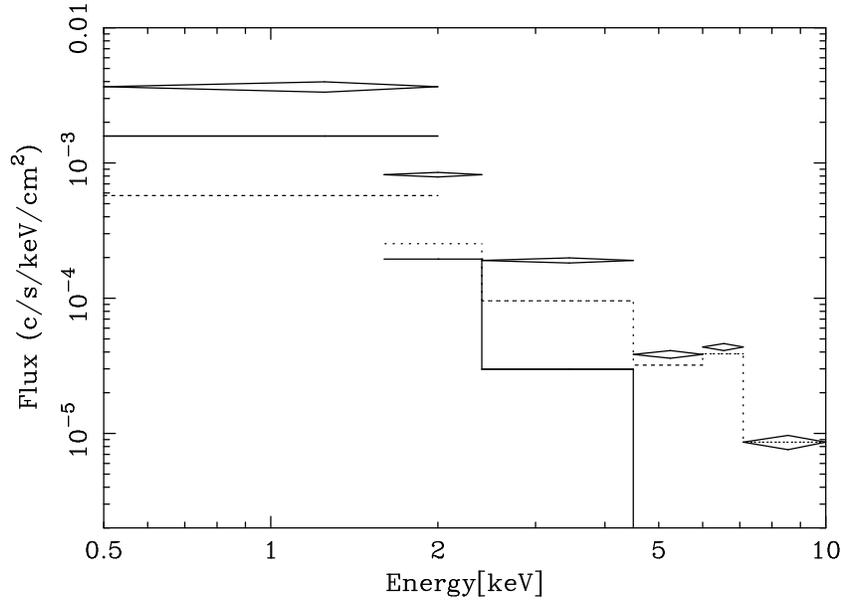,width=12.0cm,angle=-90,clip=}}
\figcaption{Deprojected energy spectrum for the central spherical
region of R$<1.5'$ (diamonds).
Energy spectra predicted by the plasma emission models of
temperature 1~keV and 4~keV are also illustrated 
with the solid and dashed histograms, respectively.
The Raymond-Smith code is used 
assuming a red-shift of 0.0104, a metallicity of 1.0 solar,
and the Galactic absorption of $8.8\times10^{20}$ cm$^{-2}$.
\label{fig:deprojected}}
\end{figure}

Now that the three-dimensional emissivity distribution, $\epsilon(R)$,
has been determined in the PSPC energy band and in the five GIS energy
bands in terms of the double-$\beta$ model,
we can calculate a ``deprojected'' 6-bin X-ray spectrum at any radius $R$. 
This is possible because the best-fit double-$\beta$ model 
has been determined in each energy band including proper normalization,
and the relative normalization between the PSPC and the GIS 
is known to an accuracy of $\sim$10\%.
In Figure~\ref{fig:deprojected}, we show the deprojected spectrum 
in the central spherical region of $R < 1'.5$ (27 kpc).
For comparison, theoretically predicted X-ray spectra from thermal plasmas
are illustrated in the same figure.

Thus, the deprojected spectrum at the cluster center
does not agree with any of the single-temperature model predictions.
Above energies of $\sim 4$~keV 
the spectrum is as hard as the $T=4$~keV model,
while the data in lower energy bands require 
contributions from much cooler (e.g. $T\sim$1~keV) plasmas.
Since the projection effects have already been removed,
we should conclude that the ICM in the central $1'.5$ region 
cannot be isothermal:
the cluster center volume must be filled with the hot ICM,
where there co-exists the cooler plasma. 
This nicely reconfirms the two-phased modeling adopted 
by FEA94 and Ikebe (1995),
who already found the hot-phase temperature of $T_{\rm h} \sim 4$~keV
and the cool-phase temperature of $T_{\rm c} \sim 1$~keV.
In contrast, the wide-band GIS data are inconsistent 
with the {\it ROSAT} deprojection results (AF94)
in which the spectrum at the cluster center is approximated
by a single-phase emission with $T \sim 1.3$~keV.
Evidently, this discrepancy arises because the {\it ROSAT} PSPC 
works only below $\sim 2$~keV.

At this stage, we arrive at the following two important specifications 
for the forward-method modeling of the Cen cluster ICM:
the double-$\beta$ model to represent spatial properties,
and the two-phase picture to express spectroscopic properties.
However, we have so far neglected the strong metallicity concentration
which was discovered previously (FEA94);
this will be taken into account in \S~6.


\section{Spectral Analysis in the Central Region}

In \S~4 we have studies the spatial properties of the X-ray emission,
and obtained significantly new findings as to the central region
of the Centaurus cluster.
However we did not fully utilize the spectral information there.
Accordingly in this section, we perform a detailed spectroscopic 
study of the central region 
by jointly analyzing spectra from the PSPC, the SIS, and the GIS.

\subsection{Derivation of the spectra in the cluster center}

\begin{figure}
\centerline{\psfig{file=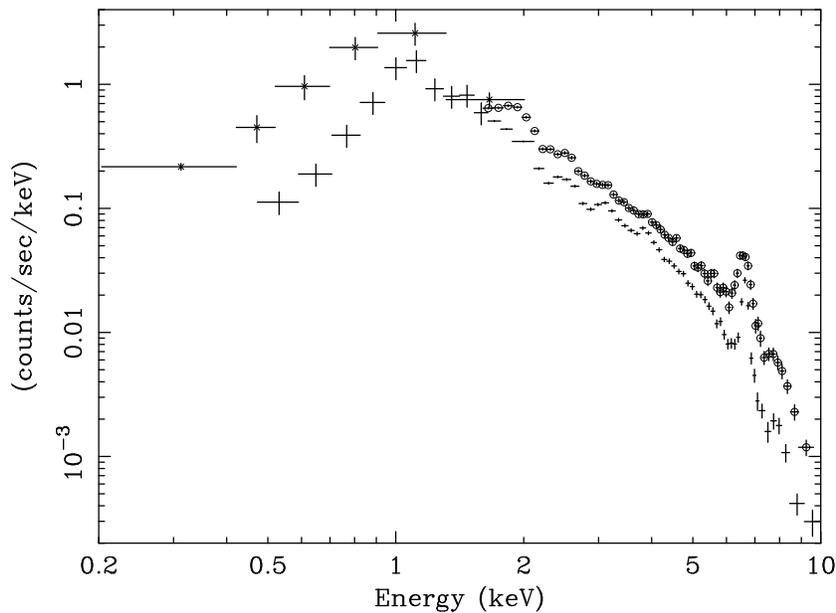,width=12.0cm,angle=-90,clip=}}
\figcaption{The PSPC (asterisks), GIS (open circles), and SIS
(crosses) spectra in the central 3 arcmin region.
\label{fig:spectra}}
\end{figure}

We accumulated energy spectra from the PSPC, GIS, and SIS,
over the central region of $3'$ radius.
The energy ranges covered by the PSPC, GIS and SIS are
0.2--2.0~keV, 1.6--10~keV, and 0.5--10~keV, respectively.
The {\it ROSAT} data used here were taken in a pointing observation 
in 1992,
and the {\it ASCA} data refer to the 1995 observation.
The PSPC background was estimated from the same observation
in the annular region of $46'-48'$ from the X-ray peak.
The data from the two GIS detectors were combined together,
as were those from the two SIS detectors.
The GIS and SIS backgrounds were produced  as described in \S~3,
utilizing the blank sky observations and the night earth data.
The background-subtracted spectra thus derived are plotted 
in Fig.~\ref{fig:spectra}.

In the previous works by AF92, FEA94, and FABT94,
individual spectra from different detectors were fitted separately.
In contrast, we fit the three spectra simultaneously,
using the XSPEC package (ver. 10.0).
This allows us to perform the highest-quality spectroscopy 
of the ICM in the central region of this important cluster.
However to obtain reliable results under rather high data statistics,
we need to consider in advance several technical issues,
which are described in the next subsection.


\subsection{Technical issues in the spectral fitting}

As a thermal plasma emission code, we use the Mewe-Kaastra model
(Mewe, Gronenschild \& van den Oord 1985;
Mewe, Lemen, \& van den Oord 1986; Kaastra 1992)
modified by Liedahl, Osterheld \& Goldstein (1995),
which is implemented in XSPEC as MEKAL model.
We use the fiducial solar abundances (relative to hydrogen in number)
given by Anders \& Grevesse (1989);
He=0.0977, C=3.63$\times$ $10^{-4}$, 
N=1.12 $\times$ $10^{-4}$,  O=8.51$\times$ $10^{-4}$,  
Ne=1.23 $\times$ $10^{-4}$, Mg=3.80$\times$ $10^{-5}$, 
Si=3.55 $\times$ $10^{-5}$, S=1.62$\times$ $10^{-5}$,  
Ar=3.63 $\times$ $10^{-6}$, Ca=2.29$\times$ $10^{-6}$, 
Fe=4.68 $\times$ $10^{-5}$, and Ni=1.78$\times$ $10^{-6}$.
The abundances of Si, S, Ar, Ca, and Fe are left free,
while that of Ni is tied to that of Fe.
Other elements are assumed to have the solar abundances.

We should here remember a serious problem of the theoretical Fe-L line
modeling, as pointed out by several authors including FABT94 and 
Arimoto et al. (1997). According to Kaastra (1997),
even the improved MEKAL model has significant uncertainties.
Therefore, we assign 20\% systematic errors to each spectral data point
in the 0.4--1.6~keV range, 
as has been employed successfully by Matsushita et al. (1997)
in analyzing the {\it ASCA} spectra of the X-ray luminous elliptical
galaxy NGC~4636.

We employ photoelectric absorption model due to the Wisconsin cross 
section (Morrison \& McCammon 1983), which is implemented in XSPEC as 
WABS model.
We fix the redshift for the PSPC data to the optical value of 0.0104,
while we leave free those for the GIS and SIS data in order to 
compensate any possible slight mismatch in their detector gains.
The relative normalizations among the three detectors are also left free.

The {\it ASCA} spectra for extended sources are generally subject
to significant flux mixing among different sky regions due to the widely 
extended PSF of the XRT.
However the spectra accumulated over the central $3'$ of the Cen cluster
are contaminated by no more than 20\% by photons from the sky regions outside,
since the brightness profile is so peaked at the center.
Furthermore, both the vignetting and the XRT PSF can be considered
as virtually constant across the $3'$-radius region.
Based on these considerations, we calculate the effective area of the 
{\it ASCA} XRT by assuming 
that a point source located at the X-ray peak is being observed.

Finally, a residual uncertainty in the SIS low energy efficiency is known
to cause a systematic over estimate of the absorption column density
(Dotani et al. 1996; Cappi et al. 1998).
Therefore, only for fitting the SIS spectrum,
we artificially apply an additional absorption to spectral models
by an equivalent hydrogen column density of $3.0\times10^{20}$ cm$^{-2}$.

\subsection{Two-temperature fits}

As already shown in \S~4 and in the previous works, 
the ICM in the central region of the Cen cluster cannot be isothermal,
but is more likely to consist of two components having different
temperatures.
Accordingly, we fit the PSPC, SIS and GIS spectra jointly
with a two-temperature thermal emission model modified by photo-electric
absorption. 
The metal abundance for each element is assumed to be common to
the two components,
and the two components are assumed to suffer the common absorption 
which is left free.
Other specifications of the fitting refer to \S 5.2.
Then, the fitting model can be expressed as WABS*(MEKAL+MEKAL) in XSPEC,
to which we shall refer as Model~1.

\begin{figure}
(a)\hspace{7.5cm}(b)
\centerline{
\psfig{file=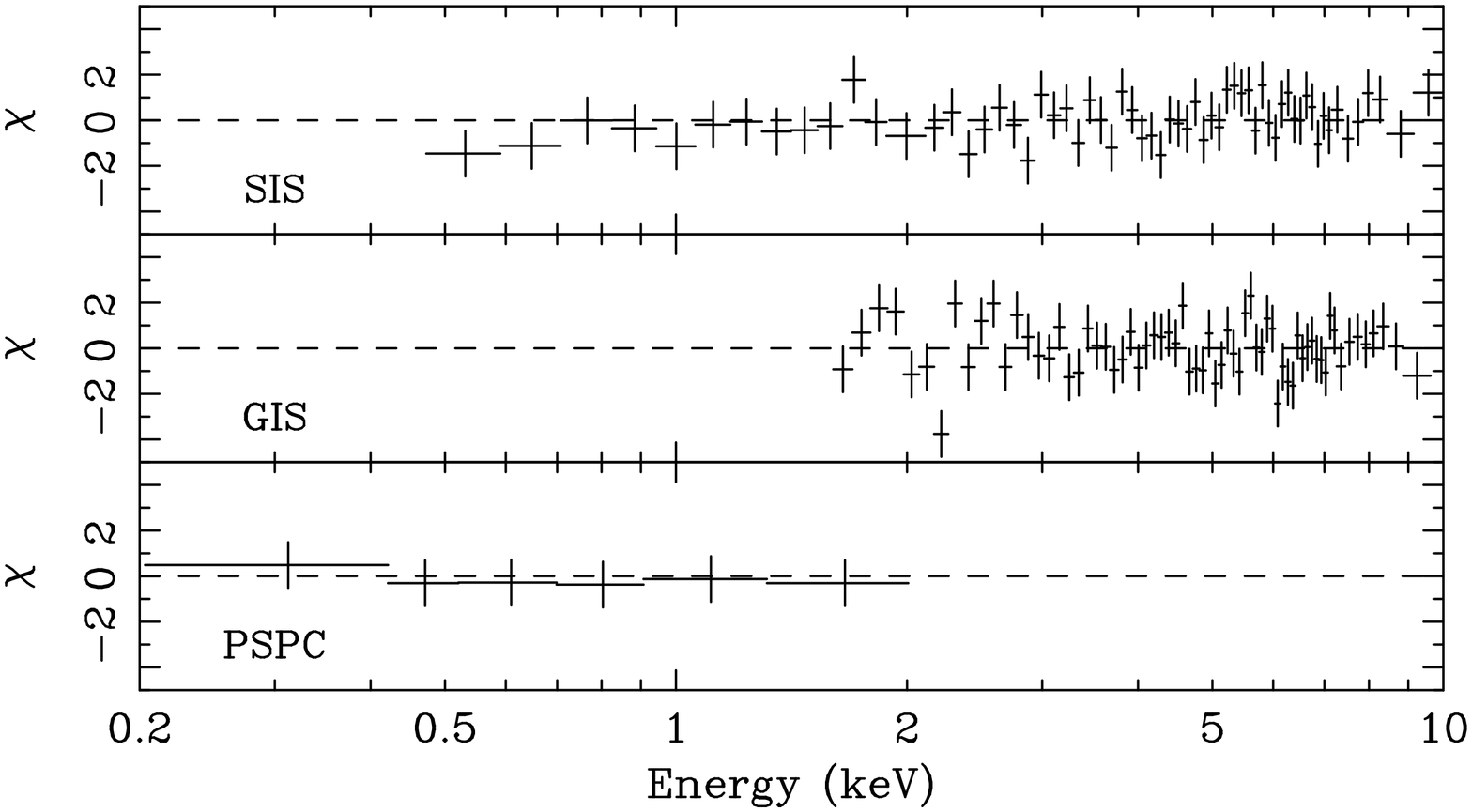,width=8.0cm,angle=-90,clip=}
\psfig{file=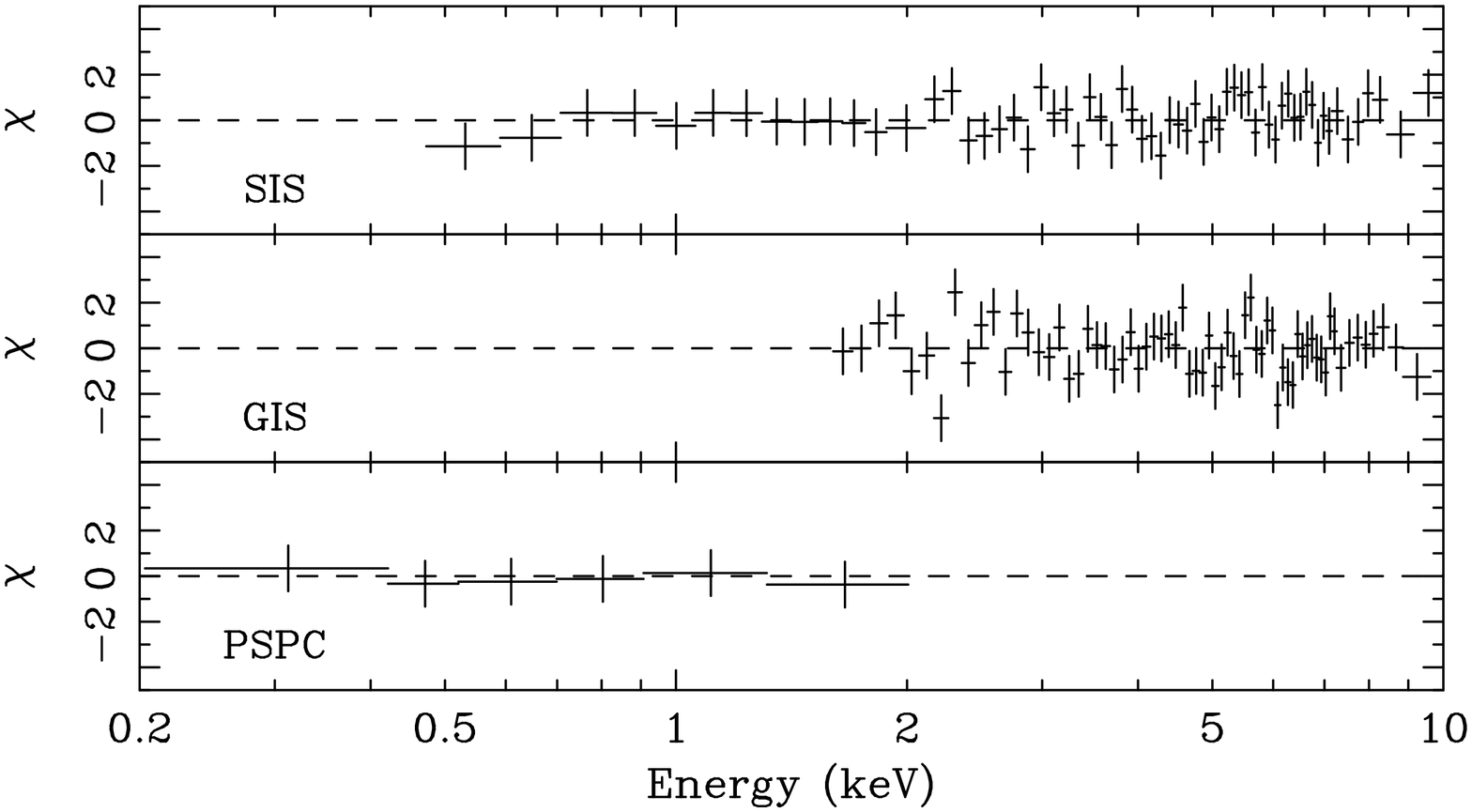,width=8.0cm,angle=-90,clip=}}

(c)

\psfig{file=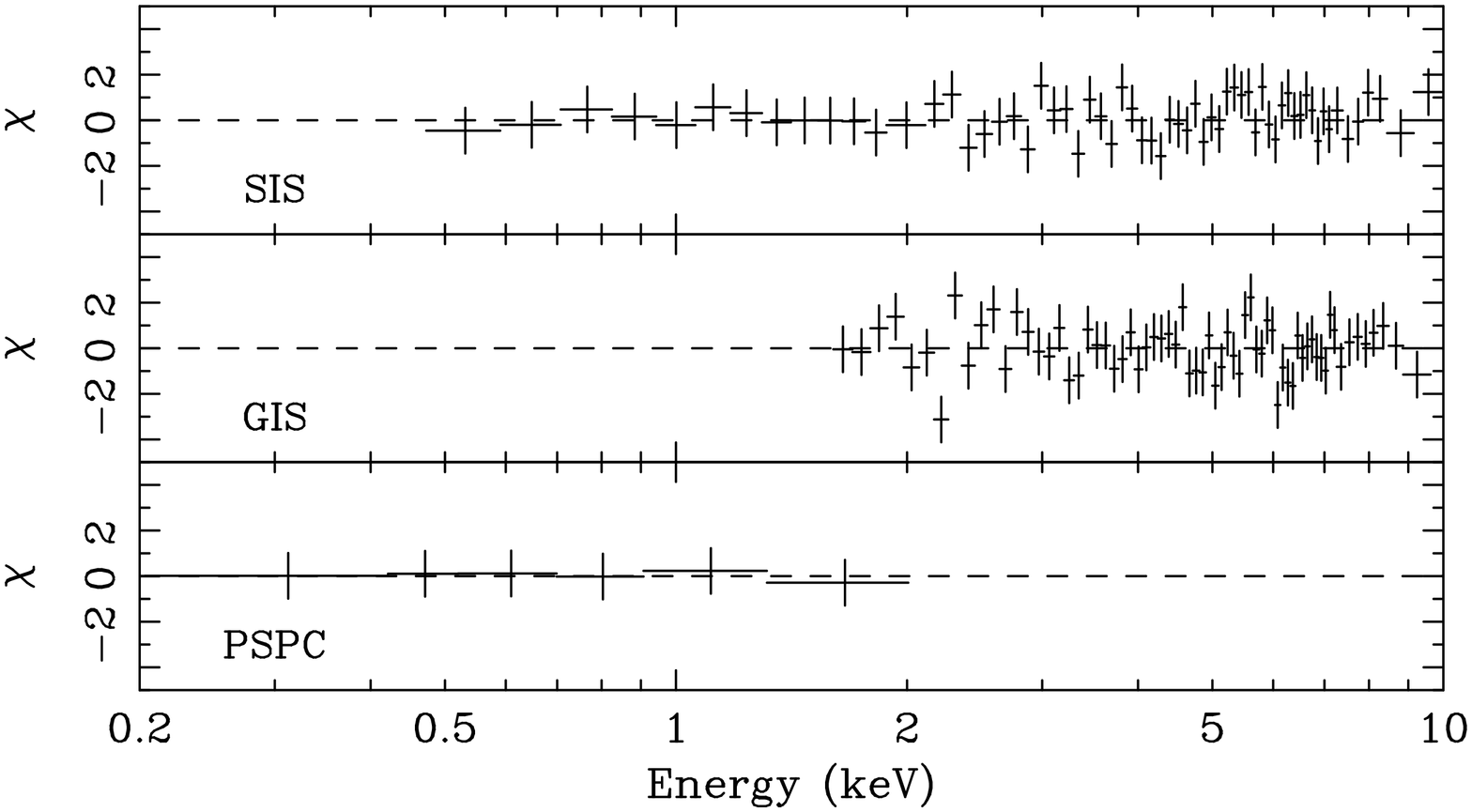,width=8.0cm,angle=-90,clip=}

(d)\hspace{7.5cm}(e)
\centerline{
\psfig{file=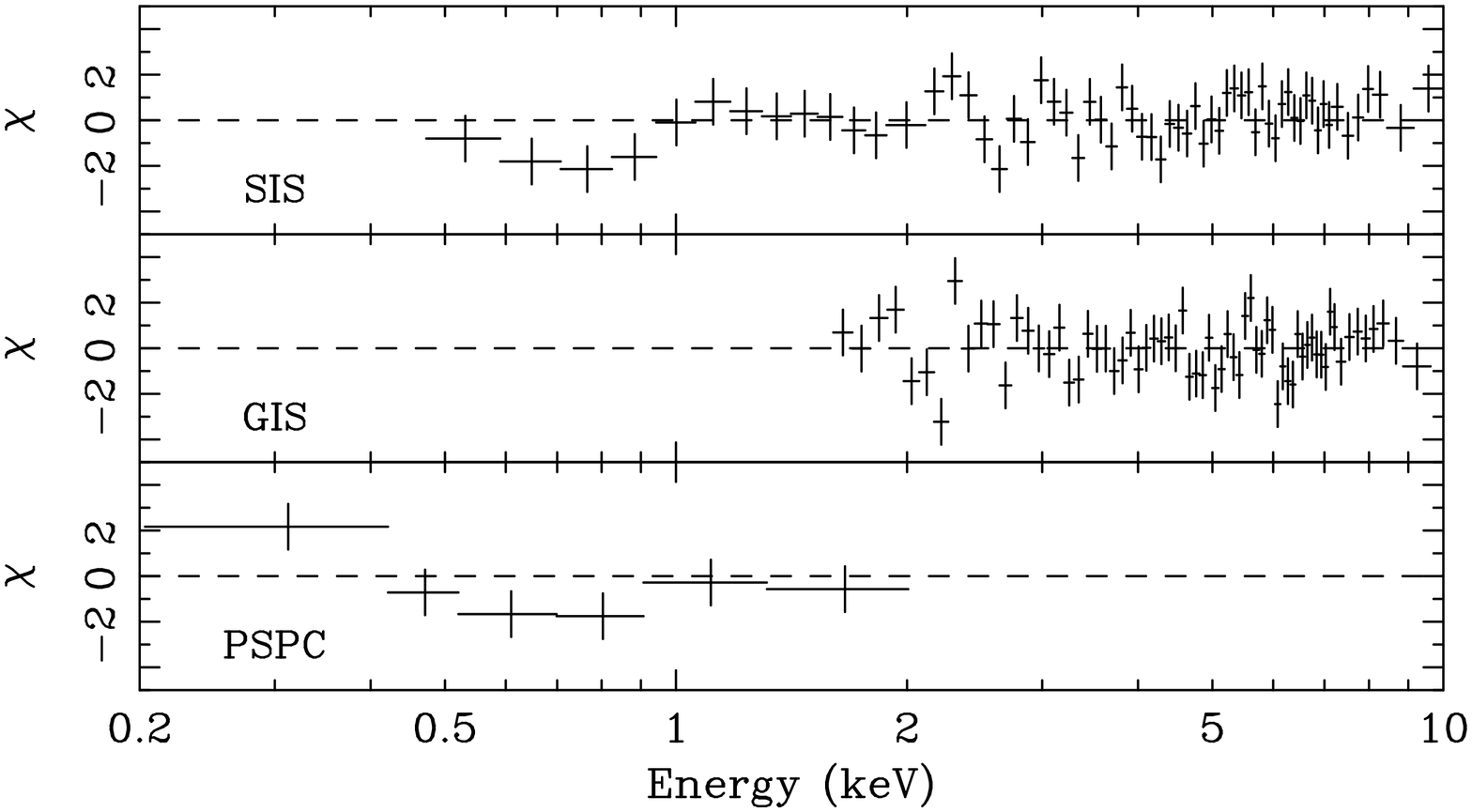,width=8.0cm,angle=-90,clip=}
\psfig{file=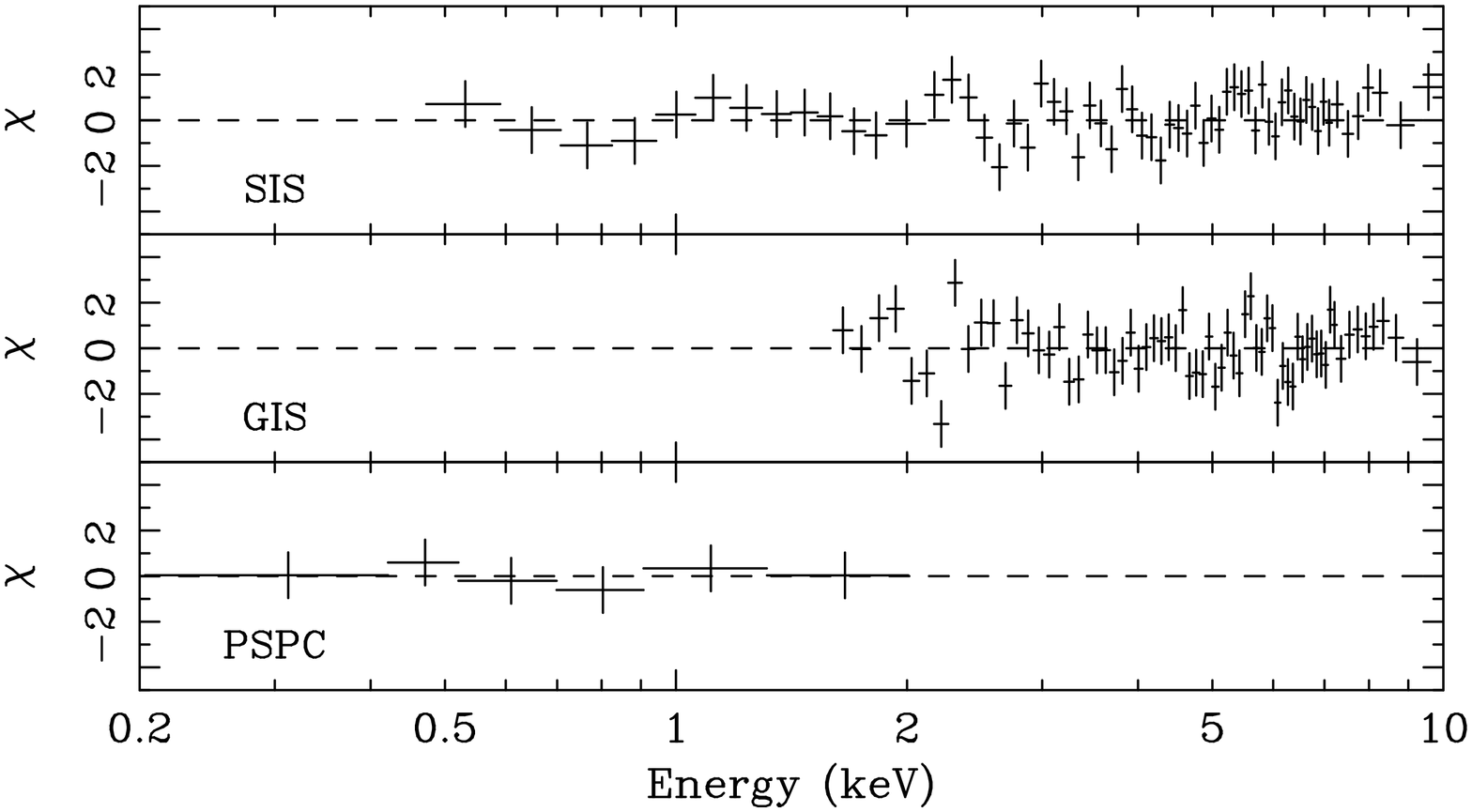,width=8.0cm,angle=-90,clip=}}
\figcaption{The fit residuals to the spectra of Figure~\ref{fig:spectra}
with (a) Model 1, (b) Model 2, (c) Model 3, (d) Model 4,
and (e) Model 5. Parameters are given in Table~\ref{tab:specfit}.
\label{fig:specfit}}
\end{figure}

This Model~1 has given an acceptable simultaneous fit,
as shown in Table~\ref{tab:specfit} and Figure~\ref{fig:specfit}a.
However, a noticeable feature remains both in the 
SIS and GIS residuals at $\sim1.7$~keV (Figure~\ref{fig:specfit}a).
Similar residuals have been noticed in the {\it ASCA} spectra 
of the Perseus and Virgo clusters (Fukazawa 1997),
and may still be caused by the uncertainty in the Fe-L complex modeling.
Therefore we have modified Model~1 
by adding one artificial narrow Gaussian line at $\sim1.7$~keV,
and refer to it as Model~2, which is expressed as WABS*(MEKAL+MEKAL+LINE).
Model~2 has given a satisfactory fit
(Table~\ref{tab:specfit}; Figure~\ref{fig:specfit}b)
with the line having a center at 1.74~keV and an equivalent width of 18~eV.
The obtained absorption column density is consistent with the Galactic
value (8.8 $\times 10^{20}$ cm$^{-2}$; Stark et al. 1992).
The redshifts for the SIS and GIS become 0.007 and 0.018, respectively.
These are consistent with the optical value of 0.0104,
because the SIS and GIS have systematic gain uncertainties of $\pm1$\%.
The best-fit metal-abundances that are common to the two components
are Si=1.6 ($\pm0.2$), S=1.7 ($\pm0.2$),
Ar=2.0 ($\pm0.4$), Ca=2.0 ($\pm0.5$), and Fe=Ni=1.00 ($\pm0.07$)
in the solar unit.
The Si and Fe abundance ratio, Si/Fe$\sim$1.6, is consistent with that
of some other clusters having the similar ICM temperature of $\sim$4~keV
(Mushotzky et al. 1996; Fukazawa et al. 1998).
When the abundance for each component was left free individually,
the Fe (=Ni) abundances of the cool and the hot component were obtained 
to be 0.8 ($^{+0.7}_{-0.5}$) and 1.00 ($^{+0.16}_{-0.08}$) solar,
respectively.
All the other metal abundances were poorly constrained in this case,
but at least have consistent values between the two components.

Previously, FABT94 reported that the two temperature model 
in which the cooler component has an significant excess absorption 
gives the best description to the SIS spectrum of the Cen cluster.
However this could be caused by the systematic error 
in the SIS low-energy response (\S 5.2), 
which was not clearly known when FABT94 analyzed the data.
In order to investigate this issue,
we applied an additional absorption to the cool component.
The model is expressed as WABS*(WABS*MEKAL+MEKAL+LINE), 
and referred to as Model~3.
The fit with Model~3 (Figure~\ref{fig:specfit}c)
reduced the minimum $\chi^2$ by only 1 from that with Model~2,
and the excess absorption remained rather unconstrained;
it is not statistically significant but can be as high as 
$2.6\times10^{21}$ cm$^{-2}$ at the cluster rest frame (90\% upper limit).
Therefore, the current datasets do not require 
an excess absorption for the cool component.

\subsection{Cooling flow fits}
In order to investigate other possible spectral models,
we replaced the cool component with a cooling flow model 
without and with the excess absorption;
they are respectively expressed as WABS*(MKCFLOW+MEKAL+LINE),
referred to as Model~4,
and WABS*(WABS*MKCFLOW+MEKAL+LINE), referred to as Model~5.
Here MKCFLOW is the cooling flow model by Mushotzky \& Szymkowiak (1988),
employing the MEKAL model as a plasma emission code,
in which the temperature is continuously distributed from the maximum
temperature $T_{\rm max}$ to the minimum temperature $T_{\rm min}$
with emissivity inversely proportional to the cooling function
at each temperature.
We fix $T_{\rm min}$ at 0.2~keV, 
while we tie $T_{\rm max}$ to the hot component temperature.

The fit results are summarized in Table~\ref{tab:specfit}
and Figure~\ref{fig:specfit}d,e.
Thus, neither Model~4 nor Model~5 gave acceptable fits,
with significantly larger values of $\chi^2$ 
than has been obtained with the two-temperature models.
The most prominent discrepancy with the Model~4 is
seen below 1~keV (Figure~\ref{fig:specfit}d).
The discrepancy is somewhat relaxed with the Model~5
(Figure~\ref{fig:specfit}d),
but it requires no Galactic absorption.
Therefore the emission is unlikely to have such a multi-temperature
structure as is predicted by the cooling flow hypothesis.

Judging from the most advanced observational information thus obtained 
through the simultaneous fits to the PSPC, SIS, and GIS spectra,
the ICM in the central region of the Cen cluster can be regarded as  
consisting of the two distinct temperature components of $\sim$1~keV
and $\sim$4~keV.
This reconfirms the main conclusion obtained previously by FEA94,
and justifies us to employ the two-temperature model rather than 
the cooling flow model.


\section{Overall Spatial and Spectral Analysis}

In order to unify the spatial results obtained in \S 4
and the spectral results derived in \S 5,
in this section we quantify the overall ICM properties
through a full forward-method analysis,
considering simultaneously the spectroscopic and spatial dimensions.
We start from a thee-dimensional model cluster,
which takes into account three important ingredients;
the two-phase temperature structure, 
the double-$\beta$ brightness distribution,
and the central metallicity increase.
We use the 1.6--10 keV GIS data and the 0.5--2.0 keV PSPC data.
Although we do not use the SIS data explicitly for 
the reason described in \S 4.2,
the analysis takes into account the spectral results obtained in \S~5
for which the SIS data have significant contributions.

\subsection{Method of analysis}

\begin{figure}
\psfig{file=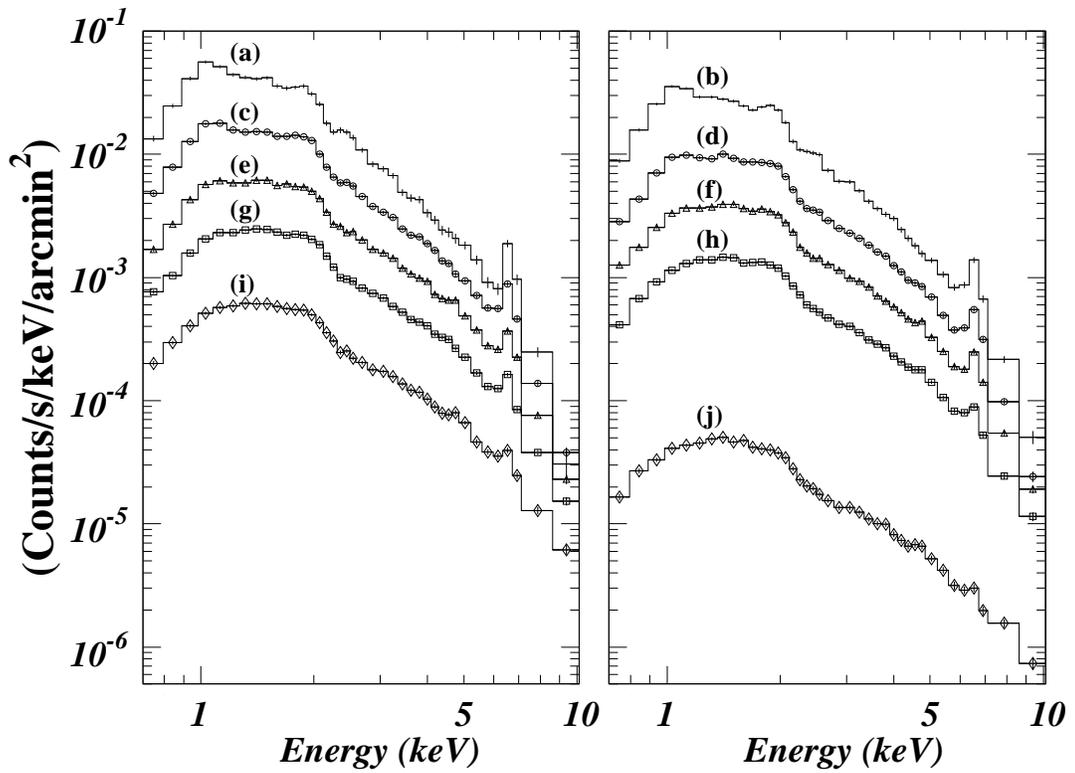,width=16.0cm,angle=0,clip=}
\figcaption{The 0.7--10 keV GIS spectra extracted from the 10 annular
regions.
Outer radii used to extract the spectra are 
(a) $1'$, (b) $2'$, (c) $3'.5$, (d) $5'$, (e) $6'.5$, 
(f) $8'.5$, (g) $11'$, (h) $14'$, (i) $18'.5$, and (j) $40'$.
\label{fig:wagirispec}}
\end{figure}

We have sorted all the available GIS data of the Cen cluster
into 10 spectra covering ten concentric annular regions, 
with outer boundaries at projected radii of 
$1'$, $2'$, $3'.5$, $5'$, $6'.5$, $8'.5$, $11'$, $14'$, $18'.5$ and $40'$.
These spectra are presented in Figure~\ref{fig:wagirispec}.
Using predictions derived from a common model cluster,
we attempt to fit jointly these 10 spectra
as well as the $r=0'-120'$ PSPC radial profile (Figure~\ref{fig:pspcrpro}).
The fitting method is essentially the same as described in \S~4,
except that the GIS data mainly constrain the spectroscopic properties
while the PSPC data strongly constrain the spatial properties.

According to the two-phase (two-temperature) modeling established in \S 5,
we express the emissivity profile in a given energy range ($E_1$ to $E_2$)
as
\begin{equation}
\epsilon(R;E_1,E_2) = 
	Q_{\rm h}(R)\cdot \Lambda(T_{\rm h},Z_{\rm h}(R);E_1,E_2)
	+ Q_{\rm c}(R)\cdot \Lambda(T_{\rm c},Z_{\rm c};E_1,E_2)~~,
\end{equation}
where $Q(R)$ is the emission measure profile, $\Lambda$ is the cooling
function, and suffices c and h denote the cool and hot phases, 
respectively.

Following \S 4.4, we represent the emission measure profile of the hot
phase,  
$Q_{\rm h}(R)$, by a double-$\beta$ model as
\begin{equation}  \label{eq:em2beta}
Q_{\rm h}(R) =  
 n_{0,1}^2 \left[1+\left(\frac{R}{R_{\rm c,1}} \right)^2
\right]^{-3\beta_1}
 + n_{0,2}^2 \left[1+\left(\frac{R}{R_{\rm c,2}} \right)^2
\right]^{-3\beta_2}~~,
\end{equation}
where $n_{0,1}$ and $n_{0,2}$ are model normalizations,
having a dimension of plasma density, to be utilized in place of 
$\epsilon_{0,1}$ and $\epsilon_{0,2}$ of equation (\ref{eq:2beta}), 
respectively.
For the cooling function $\Lambda$, we employ MEKAL model
as has been used in the previous section.
The hot-phase temperature $T_{\rm h}$ is assumed to be spatially constant,
but left free to vary.
We model the hot-phase metallicity profile $Z_{\rm h}(R)$ as
\begin{equation} \label{eq:metallicity}
   Z_{\rm h}(R) = \Delta Z \exp\left( -\frac{R}{R_{\rm Z}} \right) + Z_0~~.
\end{equation}
Here $Z_0$ is the outer-region metallicity,
$\Delta Z$ is the central metallicity increment, 
and $R_{\rm Z} >0$ is the angular scale of the metallicity gradient,
all of which are free parameters to be determined.
We assume $Z_{\rm h}(R)$ to be common to the two $\beta$-components 
of equation (\ref{eq:em2beta}).

As to the cool phase, 
we model its radial emission measure profile, $Q_{\rm c}(R)$, 
with a single-$\beta$ model involving three free parameters, as
\begin{equation} \label{eq:em1beta}
Q_{\rm c}(R) = n_{\rm 0,c}^2\ 
\left[ 1 + \left(\frac{R}{R_{\rm c,c}}\right)^2 \right]^{-3\beta_{\rm c}}~~.
\end{equation}
The cool-phase spectrum, $\Lambda(T_{\rm c},Z_{\rm c};E_1,E_2)$,
is assumed to be spatially constant,
with $T_{\rm c}$ and $Z_{\rm c}$ 
being the temperature and metallicity, respectively.
We also introduce a cut-off radius, $R_{\rm cut}$,
beyond which there no longer exists the cool phase. From 
the {\it ROSAT} results (AF94), we fix $R_{\rm cut}=6'.0$.

The cluster model introduced here involves many free parameters.
The hot component involves at least 10 free parameters, 
$T_{\rm h}$, $n_{0,1}$, $n_{0,2}$, 
$R_{\rm c,1}$, $R_{\rm c,2}$, $\beta_1$, $\beta_2$,
$Z_0$, $\Delta Z$, and $R_{\rm Z}$.
If the abundance ratios among the different elements are allowed to vary,
the number of free parameters becomes still larger.
For the cool component there are at least 5 parameters,
$T_{\rm c}$, $Z_{\rm c}$, $n_{\rm 0,c}$, $R_{\rm c,c}$, and $\beta_c$,  
and even more if allowing variable abundance ratios.
In addition, we leave free the relative normalization 
between the PSPC and the GIS,
as well as the PSPC background which is assumed to be radius independent.
Thus, the number of model free parameters becomes at least 17.

\subsection{Results of the analysis}

In actually fitting the 10 GIS spectra and one PSPC radial profile
simultaneously with the cluster model described in \S 6.1,
it is not practical to leave all the parameters free.
Accordingly, we fix the chemical composition of the hot phase
to obey the abundance ratios obtained with Model~2 in \S~5.3,
and model the iron abundance profile with equation (\ref{eq:metallicity}),
so that abundances of all the other heavy elements vary with $R$
in proportion to that of iron.
This leaves the hot phase with 10 free parameters.
We also fix the temperature and various abundances of the cool phase
($T_{\rm c}$, $Z_{\rm c}$), and the hydrogen column density,
all to the values derived with Model~2 in \S~5.3.
This also allows us to best utilize the SIS information, 
which is not incorporated directly here.
Consequently, we have three free parameters specifying the cool component,
$n_{\rm 0,c}$, $R_{\rm c,c}$, and $\beta_c$.
Adding the relative PSPC/GIS normalization and the PSPC background,
the number of free parameters becomes 15.

\begin{figure}
\centerline{
\hspace{3.0cm}\psfig{file=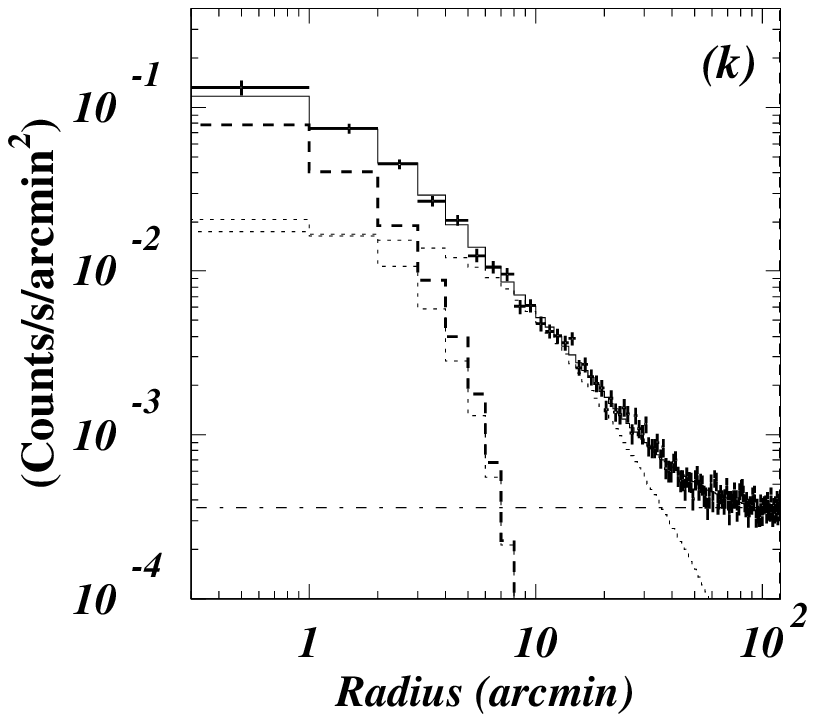,width=6.0cm,angle=0,clip=}}
\vspace{-4.5cm}
\psfig{file=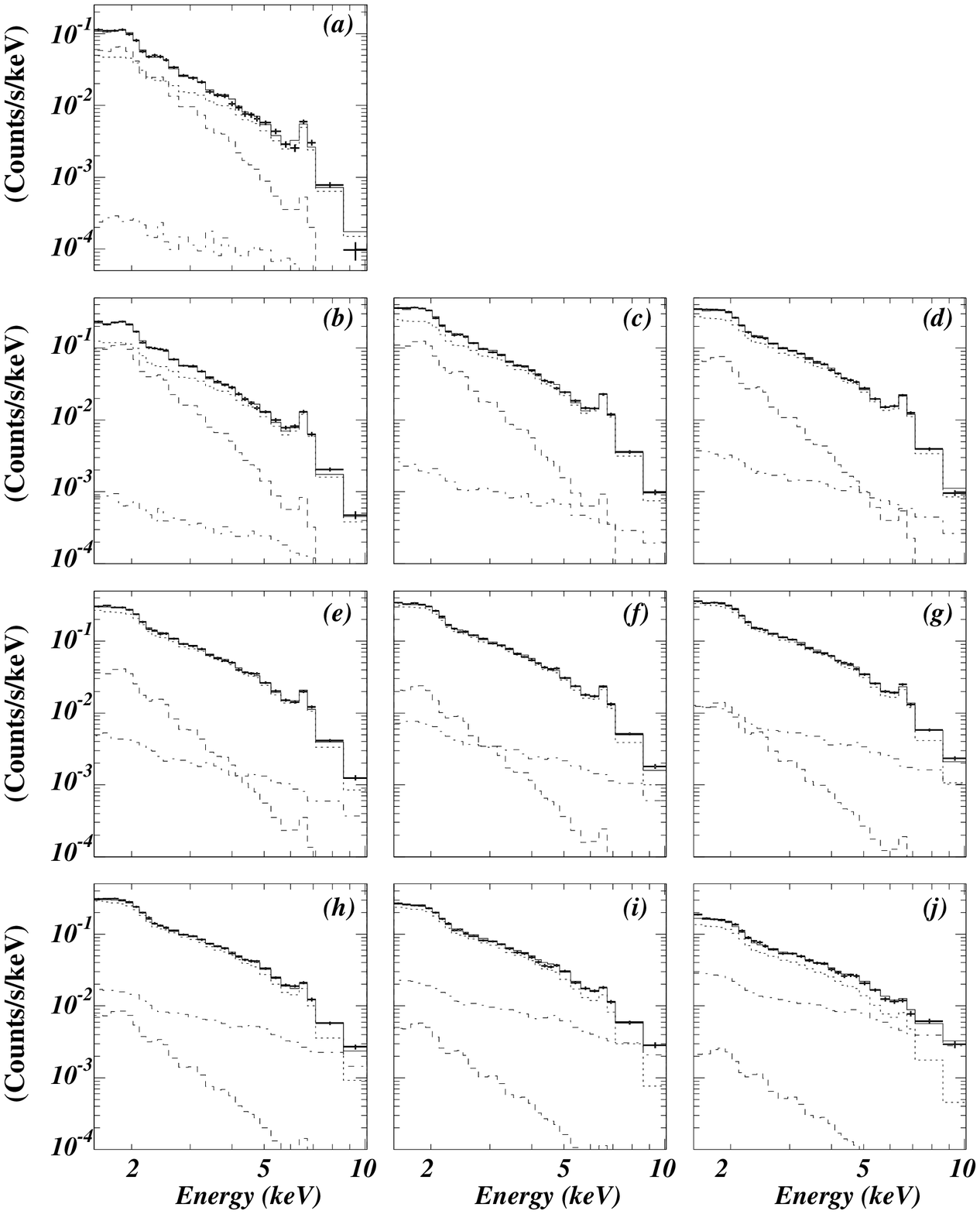,width=15.0cm,angle=0,clip=}
\figcaption{The simultaneous fit to the 10 GIS annular spectra
(a-j: the same as figure~\ref{fig:wagirispec})
and the PSPC radial profile (k) with the two-phase model.
The energy ranges used for the fit are 0.5--2~keV
and 1.6--10~keV for the PSPC and GIS data, respectively.
Crosses show the data, and solid histograms show
the fitted models convolved with the responses of 
the X-ray telescopes and the detectors.
Three components in the model,
i.e. the hot component, the cool component, and the background,
are also illustrated with dotted histograms, dashed histograms,
and dot-dashed histograms, respectively.
In panel (k), the two $\beta$ components constituting the hot-phase
emission are drawn separately.
\label{fig:2phasefit}}
\end{figure}

\begin{figure}

\hspace{1cm}(a)\hspace{7.7cm}(b)

\vspace{-10mm}
\centerline{
\psfig{file=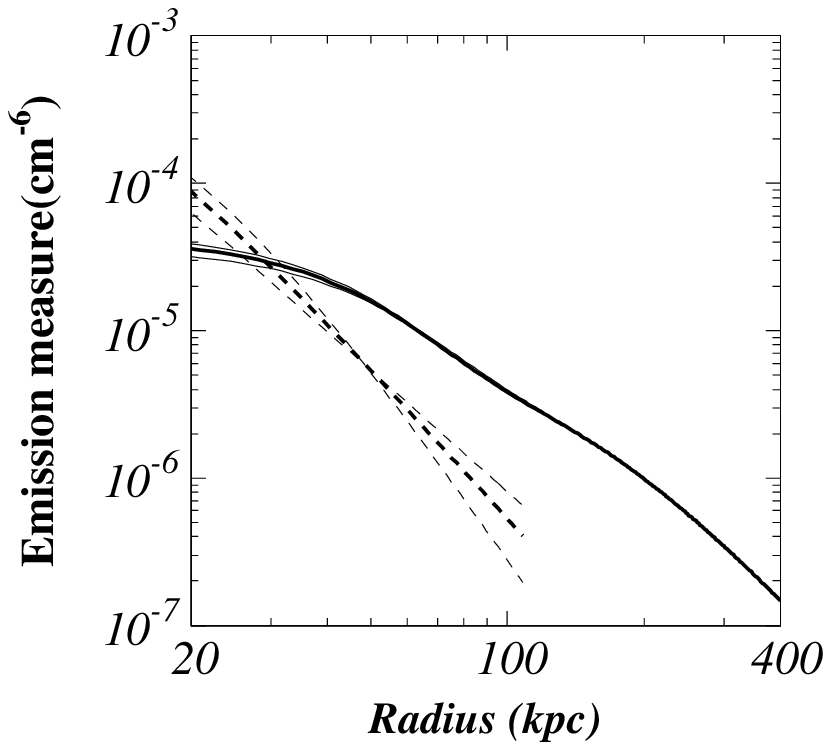,width=8.0cm,angle=0,clip=}
\psfig{file=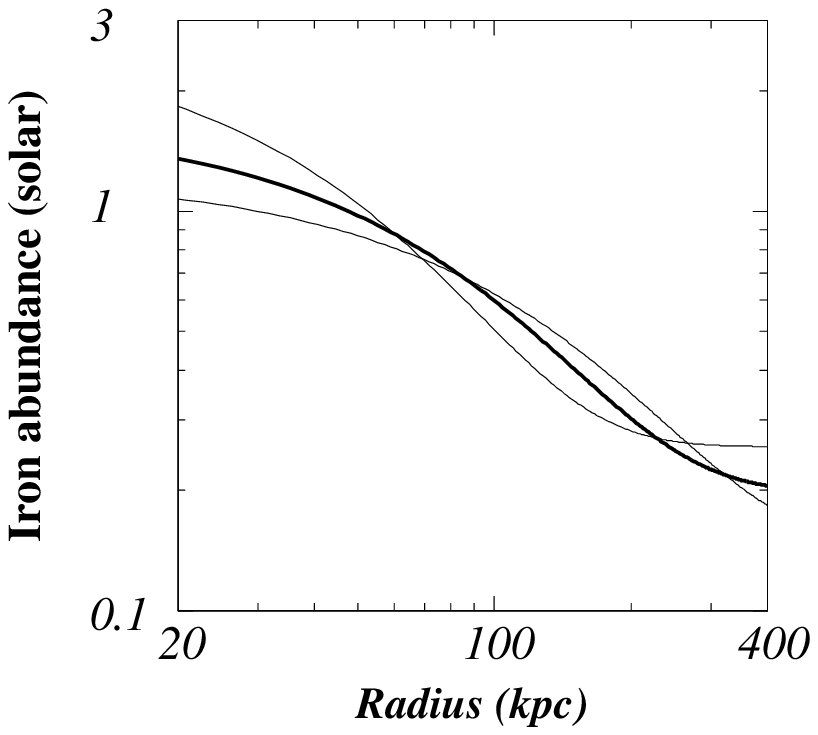,width=8.0cm,angle=0,clip=}}
\figcaption{(a) The emission measure profiles of 
the hot (solid lines) and cool (dashed lines) component,
and (b) the iron abundance profile derived from the two-phase model fit.
Thick lines show the best fit models, while thin
lines indicate extreme cases that produce
a $\Delta \chi^2$ = 2.706 corresponding to the 90\% confidence limit.
\label{fig:2phasefitmodel}}
\end{figure}

By changing the 15 parameters and minimizing the chi-square,
we search for the model which gives acceptable simultaneous fits 
to the 10 ring-sorted GIS spectra plus the PSPC radial profile.
Then, as shown in Figure~\ref{fig:2phasefit}, 
an acceptable best-fit model has been obtained with $\chi^2/\nu = 377/405$.
The best-fit parameters are summarized in Table~\ref{tab:ringspecfit},
and the model emission measures of the hot and cool components
and the hot-component iron-abundance profile are illustrated 
in Fig.~\ref{fig:2phasefitmodel}.

The obtained hot-component temperature of $T_{\rm h}=3.9$~keV is somewhat
lower than the value derived from the central $3'$ region 
(Table~\ref{tab:specfit}),
possibly indicating a global outward temperature decrease in the hot
component as has been observed in many clusters (Markevitch et al. 1998).
However the difference is only 10\%, so that we can safely regard 
the hot phase as being approximately isothermal.
The wider-component parameters,
$R_{\rm c,1}=7.'5$ $^{+2.1}_{-0.8}$ and $\beta_1=0.57$ $^{+0.04}_{-0.02}$,
are close to those derived in \S 4.3 using the $r=6'-120'$ {\it ROSAT}
data.
As shown in Table~\ref{tab:ringspecfit},
the narrower-component parameters are poorly constrained.
To assess the relative spatial scales between the two $\beta$ components,
we again performed the fit, forcing the narrower-component $\beta$ value,
$\beta_2$, to be equal to that of the wider-component, $\beta_1$.
The derived core radii are $R_{\rm c,1}=9'.6$ and $R_{\rm c,2}=2'.3$
for the wider and narrower $\beta$ components, respectively,
with the common $\beta$ of 0.59
and the relative normalizations of 
$n_{0,1}$:$n_{0,2}$:$n_{0,c}$=1:3.9:16.4,
yielding a chi-square of $\chi^2/\nu = 380/406$.
This agrees with the results in the double-$\beta$ fits 
(Table~\ref{tab:rprofit}).
Also the values of $Z_0=0.2$, $\Delta Z=1.5$, and $R_{\rm Z}=4.'1$
are not much different from those found previously (e.g. FEA94; Ikebe
1995).

As to the cool-phase emission measure profile, $Q_{\rm c}(R)$, 
the best-fit solution yields $R_{\rm c,c}=0.'5$ $^{+0.5}_{-0.3}$
and $\beta_{\rm c}=0.57$ $^{+0.17}_{-0.11}$.
Therefore $Q_{\rm c}(R)$ is inferred to have a slightly smaller 
core radius than the narrower $\beta$-component describing $Q_{\rm h}(R)$.
In other words, the cool emission is somewhat more concentrated 
to the cluster center than the hard-band central excess emission.

Based on these results, we conclude
that the two-phase model has been accepted 
by the {\it ASCA} and {\it ROSAT} data. 
Furthermore, the obtained parameters of the two-phase model are 
generally consistent with those obtained previously 
by FEA94, AF94, and Ikebe (1995).


\subsection{Filling factor and the ICM mass distributions}

\begin{figure}
\centerline{
\psfig{file=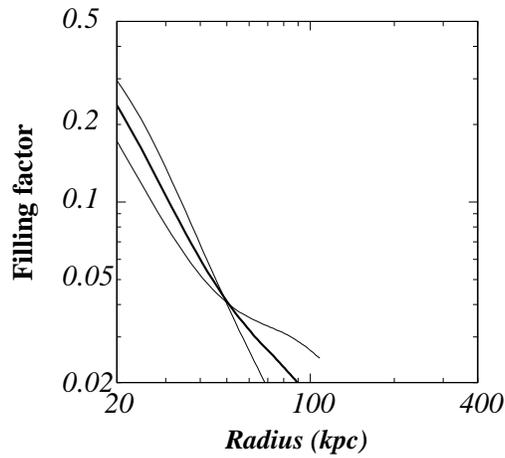,width=8.0cm,angle=0,clip=}}
\figcaption{Radial profiles of the filling factor of the cool phase.
The thick line and the thin lines represent the same cases as
described in Fig.~\ref{fig:2phasefitmodel}.
\label{fig:fillpro}}
\end{figure}

In order to better illustrate the two-phase solution obtained in \S~6.2,
we introduce the volume filling factor $f(R)$ of the cool phase.
We assume, after FEA94 and Ikebe (1995),
that a local pressure balance is achieved between the two phases, i.e.,
\begin{equation} \label{eq:pbal}
n_{\rm c}(R) \ T_{\rm c} = n_{\rm h}(R) \ T_{\rm h} \;, 
\end{equation}
and the emission measure of the two phases can be expressed as
\begin{equation} \label{eq:emdens}
     Q_{\rm h}(R)= n_{\rm h}(R)^2 \left\{1-f(R)\right\},~~~~\\
     Q_{\rm c}(R)= n_{\rm c}(R)^2 f(R) ~~.
\end{equation}
Here $n_{\rm h}(R)$ and $n_{\rm c}(R)$ are 
densities of the hot and cool phase, respectively,
and $f(R)$ is the volume filling factor of the cool phase.
These two equations yield 
\begin{equation} \label{eq:fill}
    f(R) = \left\{ 1 + \left(\frac{T_{\rm h}}{T_{\rm c}}\right) ^2 
    \frac{Q_{\rm h}(R)}{Q_{\rm c}(R)} \right\}^{-1} \; . 
\end{equation}
We may then substitute the best-fit double-$\beta$ model into $Q_{\rm
h}(R)$,
and the best-fit single-$\beta$ model into $Q_{\rm c}(R)$, to obtain
$f(R)$.
The result is shown in Figure~\ref{fig:fillpro}.
Within $R = 60$ kpc, the average filling factor of 
the cool phase becomes 6--8\%.

\begin{figure}
\centerline{
\psfig{file=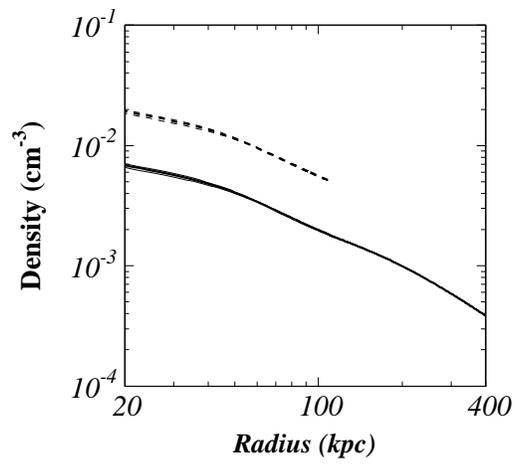,width=8.0cm,angle=0,clip=}}
\figcaption{Radial profiles of the electron densities 
of the hot (solid lines) and cool components (dashed lines).
The thick line and the thin lines represent the same cases as
described in Fig.~\ref{fig:2phasefitmodel}.
\label{fig:denspro}}
\end{figure}

We may now substitute $f(R)$ back into equation (\ref{eq:emdens}),
and convert the emission measure distributions into 
radial density distributions of the hot and cool components,
$n_{\rm h}$ and $n_{\rm c}$, respectively.
The calculated ICM density distributions are shown in
Figure~\ref{fig:denspro}.
We can further calculate the mass profiles of the cool 
and hot ICM components as
$\int 2 \mu m_{\rm p} n_{\rm c} f dV$ and 
$\int 2 \mu m_{\rm p} n_{\rm h} (1-f) dV$ ,
respectively,
where $\mu$ is the mean molecular weight which is set to be 0.6,
and $m_{\rm p}$ is the proton mass.
We show the calculated ICM mass distributions in
Figure~\ref{fig:masspro}a,
together with the spherically integrated stellar mass distribution
derived assuming the mass-to-light ratio of 8 from 
the optical brightness data of Schombert (1987) and Dickens et al. (1986).
The integrated iron mass profiles in the cool and hot phase
were also calculated and illustrated in Fig.~\ref{fig:masspro}b.


\subsection{The gravitating mass distribution}

Using the best-fit two-phase model,
we also calculated the radial profile 
of the total gravitating mass, $M_{\rm tot}$, as 
\begin{equation} \label{eq:grav}
 M_{\rm tot} (R) = 
	- \frac{R^2}{\rho_{\rm g} G} \frac{{\rm d}P}{{\rm d} R}\; ,
\end{equation}
where $P$ is the ICM pressure, $\rho_g$ is the mass density of ICM,
and $G$ is the constant of gravity.
This equation is based on the standard formula for 
an isothermal ICM in hydrostatic equilibrium (e.g. Sarazin 1988).
The pressure $P$ is given as 
$P$ = $2n_{\rm c}k_{\rm B}T_{\rm c}$ = $2n_{\rm h}k_{\rm B}T_{\rm h}$
($k_{\rm B}$ is the Boltzmann constant),
while the gas mass density $\rho_{\rm g}$ can be expressed as
\begin{equation} \label{eq:meanrho}
\rho_{\rm g} = 2 \mu m_{\rm p}\{fn_{\rm c}+(1-f)n_{\rm h}\}~~.
\end{equation}

The calculated $M_{\rm tot} (R)$ is shown in Figure~\ref{fig:masspro}c.
There, we also present dark-matter distribution, $M_{\rm DM} (R)$, 
derived by subtracting the ICM mass and the stellar mass from 
$M_{\rm tot}(R)$.
$M_{\rm tot}(R)$ agrees quite well with the result by AF94,
except in the central region where the two-phases are coexisting.
Figure~\ref{fig:masspro}d gives the $R$-dependence of the baryon
fraction,
i.e. the stellar mass plus ICM mass normalized to the total mass.

In Figure~\ref{fig:masspro}c, both $M_{\rm tot} (R)$ and $M_{\rm DM}(R)$
exhibit a shoulder-like feature at $R \sim 80$ kpc.
This feature arises through the following two effects.
One is that the emissivity profile requires two $\beta$ components 
even in the high energy range, as shown in \S~4.4 and \S~6.2.
Therefore, the gravitational potential must exhibit an additional
drop within $\sim 80$ kpc, in order to confine the excess ICM.
The other is that the central excess in $n_{\rm h}$ 
(the gas density of the hot component, hence pressure) 
is in fact still more prominent than 
that seen in $\epsilon_{\rm h}(R)$ or $Q_{\rm h}(R)$,
because of the $(1-f)$ factor in equation (\ref{eq:emdens}).
In other words, the central excess is expected to be 
less prominent in the hard-band emissivity (or emission measure)
than in the hot-phase pressure, because the hot phase is partly displaced
by the cool phase near the cluster center.
A further discussion is given in \S~7.2.


\section{Summary and Discussion}

\subsection{Summary of results}

We have analyzed the most advanced X-ray data of the Centaurus cluster
obtained with the {\it ROSAT} PSPC, the {\it ASCA} GIS,
and the {\it ASCA} SIS.
In \S~4, we showed that
the radial count-rate profiles in several broad energy bands
obtained with the GIS and the PSPC
are described adequately with the double-$\beta$ models
(Figure~\ref{fig:pspcrpro}c \& \ref{fig:gisrpro}c,
Table~\ref{tab:rprofit}).
This means that the central excess brightness within $R \sim 6'$
that has been known in soft X-rays
is present over the entire {\it ASCA} energy range up to 10~keV,
although the excess fraction is lower in higher energies
(Figure~\ref{fig:rce}).
The ``deprojected'' six-band spectrum at the center 
(Figure~\ref{fig:deprojected}) is clearly non-isothermal,
indicating a multi-phase property.
This implies that the single-phase approximation
that was often used to describe soft X-ray results on clusters
is no longer valid at least for the Cen cluster.

In \S5, we jointly analyzed the X-ray energy spectra taken with
the PSPC, GIS, and SIS, and showed that
the two-temperature ($\sim1$~keV and $\sim4$~keV) thermal-emission model 
modified by the Galactic absorption gives the best description of the
spectra, reconfirming the main conclusion of the previous work by FEA94.
The spectra did not require significant excess absorption 
to the cool component.
The cooling flow model, 
either with or without the excess absorption,
did not give acceptable fits to the spectra.

Combining the spatial and spectral properties 
derived respectively in \S4 and \S5,
we have obtained in \S6 the overall ICM model by simultaneously fitting 
the ring-sorted GIS spectra and the PSPC radial profile.
The best-fit solution indicates that,
within $\sim 6'$ ($\sim 100$ kpc) of the cD galaxy NGC~4696, 
the ICM can be regarded as an unresolved mixture of two disjoint phases;
a hot phase with $T_{\rm h} \sim 4$ keV
of which the emission measure profile can be expressed by
a double-$\beta$ model with core radii $\sim$175~kpc and $\sim$43~kpc,
and a cool phase with $T_{\rm c} \sim 1$ keV
of which the emission measure profile is described 
by a single-$\beta$ model having a small core radius of $\sim$10~kpc.
The dramatic metallicity increase near the center, 
first revealed by FEA94, has also been confirmed.
We summarize the characteristics of the two phases in 
Table~\ref{tab:summary}.

These results consistently indicate that the central region 
within $R \sim 3'$ (50 kpc)  of the Cen cluster 
is a site of three characteristic phenomena.
One is the hard X-ray excess emissivity,
revealed in \S~4.3 and reconfirmed in \S~6.2.
Another is the co-existence of the hot and cool phases, 
with the cool-phase filling factor behaving as shown 
in Figure~\ref{fig:fillpro}.
The other is the high metallicity in both phases, 
as shown in \S~5.3 and Figure~\ref{fig:2phasefitmodel}.


\subsection{Dark matter distribution}
\begin{figure}
\hspace{1cm}(a)\hspace{7.7cm}(b)

\vspace{-10mm}
\centerline{
\psfig{file=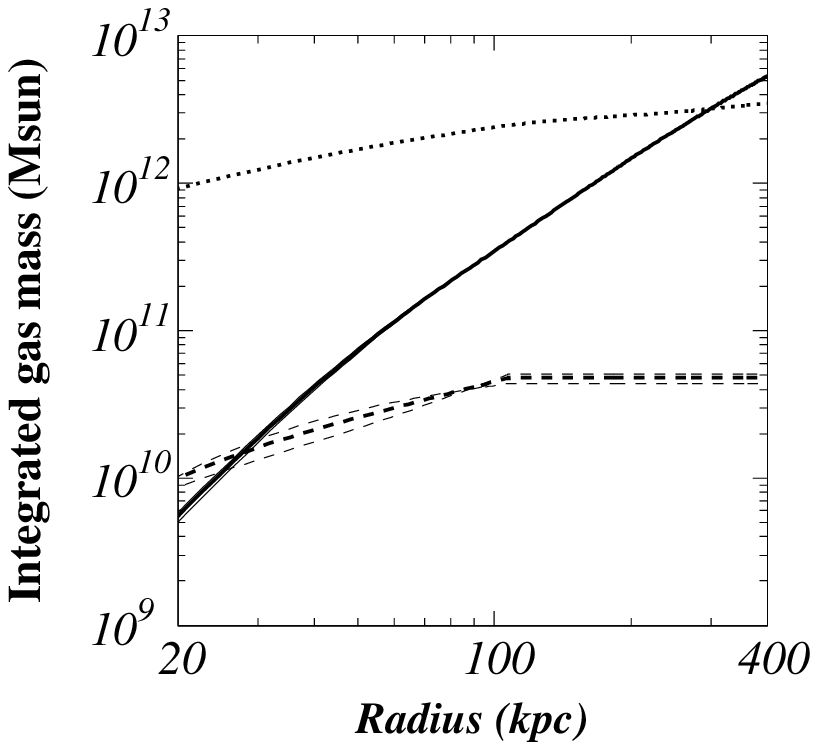,width=8.0cm,angle=0,clip=}
\psfig{file=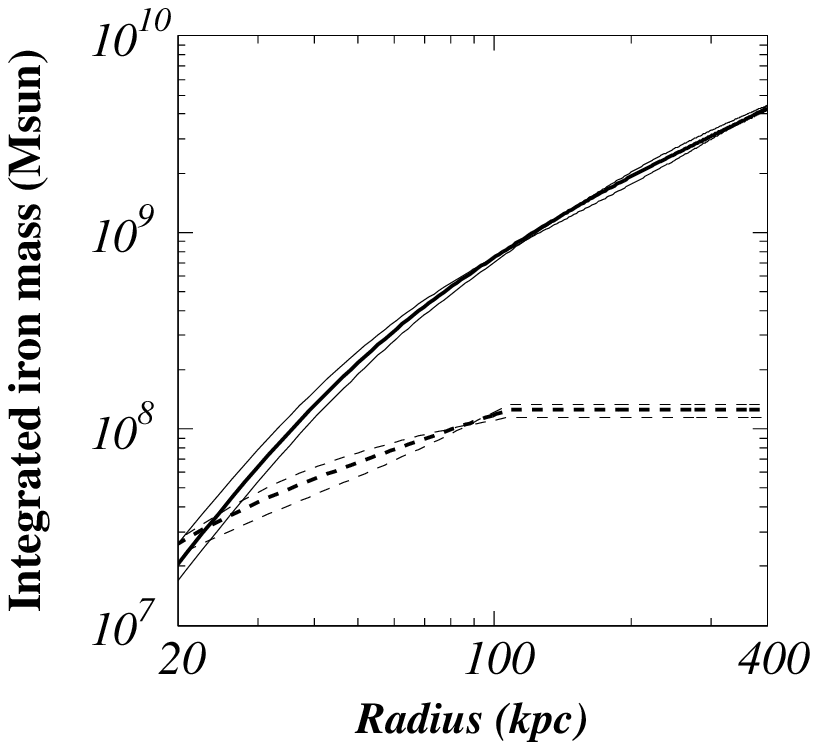,width=8.0cm,angle=0,clip=}}
\hspace{1cm}(c)\hspace{7.7cm}(d)

\vspace{-10mm}
\centerline{
\psfig{file=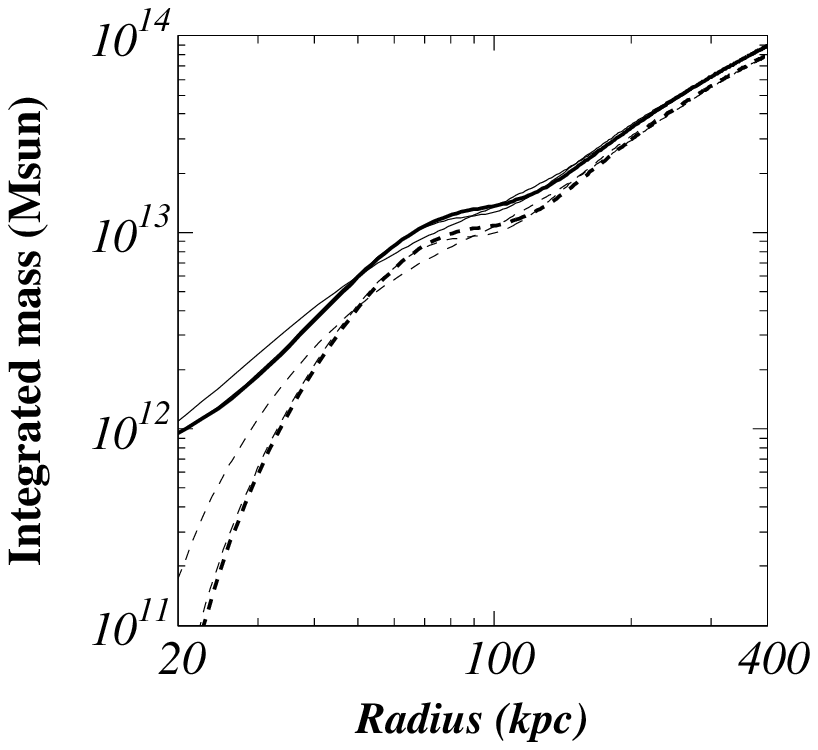,width=8.0cm,angle=0,clip=}
\psfig{file=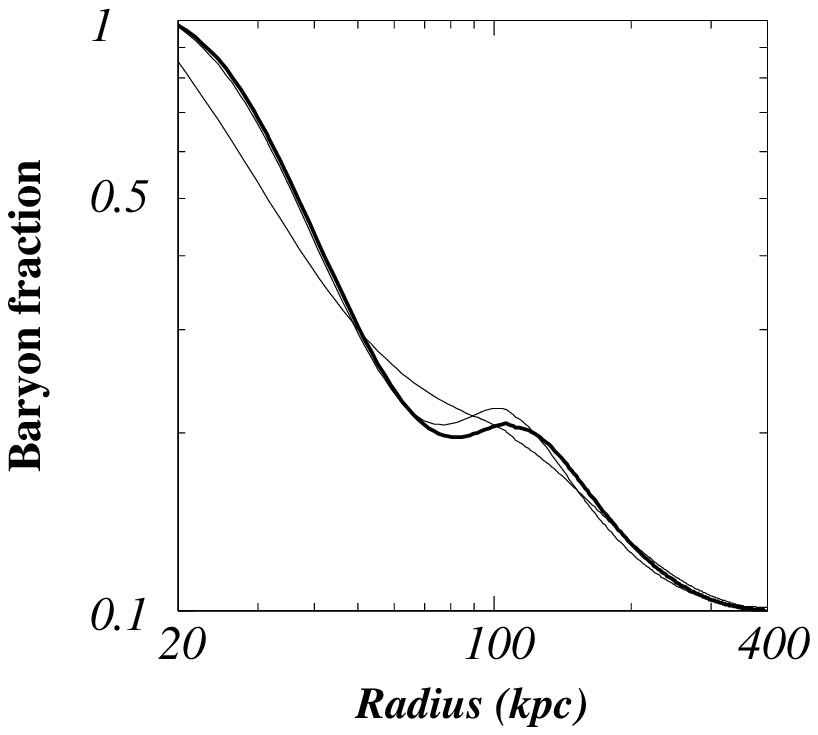,width=8.0cm,angle=0,clip=}}
\figcaption{(a) The integrated radial mass profiles of the ICM
in the hot (solid lines) and cool (dashed lines) phases, 
shown together with the stellar-component mass (dotted line).
(b) The integrated iron mass in the hot (solid lines) and 
cool (dashed lines) phases.
(c) The integrated mass of the total gravitating matter (solid lines)
and the dark matter (dashed lines).
(d) The baryon fraction.
In the all panels, the thick line shows the best-fit model,
while thin lines show extreme cases within 90\% ($\Delta \chi^2$=2.706)
confidence limits.
\label{fig:masspro}}
\end{figure}

The two-phase modeling of the Cen cluster ICM has yielded the integrated
mass distributions as shown in Figure~\ref{fig:masspro}c (\S 6.4).
At $R \sim 80$ kpc, the total mass curve
exhibits a characteristic bending feature.
Although details of this feature may depend on the ICM  modeling,
the deviation from a King-type profile is a robust result,
because it directly results from the central hard X-ray excess
above the single-$\beta$ model (\S~4.3).
In other words, the potential profile of the Centaurus cluster becomes
deeper at the center than a King-type one that has a flat core,
and this deepening causes the central excess brightness in hard X-rays.
In soft X-rays, the central excess emission is enhanced by the cool phase.

As seen from Figure~\ref{fig:masspro}d, this central mass excess
is significantly contributed by the stellar mass in the cD galaxy.
Nevertheless, even after subtracting the stellar mass contribution,
the dark matter profile (Figure~\ref{fig:masspro}c) still appears to
show an inflection point at $R \sim 80$ kpc as the total mass profile.
Therefore, the dark matter distribution itself is
inferred to deviate from a King-type profile.

Such a central excess in the dark-matter distribution has been observed
in the Fornax cluster (Makishima 1995; Ikebe 1995; Ikebe et al. 1996),
Abell 1795 (Xu et al. 1998),
the elliptical galaxy NGC~4636 (Matsushita et al. 1998),
and several galaxy groups (Mulchaey \& Zabludoff 1998).
These dark matter profiles have been interpreted
as representing a galaxy/cluster hierarchy;
the dark matter is clustering on two separate spatial scales
corresponding to the central galaxy and the entire cluster.
The effect is above all evident in the Fornax cluster (Ikebe et al. 1996),
since the X-ray peak associated with the cD galaxy, NGC~1399,
is offset by 40--50 kpc from the center of the cluster emission.
Furthermore in NGC~4636, the presence of two separate spatial scales has
been confirmed in a model independent manner by Matsushita et al. (1998).
Then, the dark matter in the Cen cluster may
also show such a dual distribution.

Alternatively, this dual structure may be an artifact caused
by our particular modeling using two $\beta$-models,
and the dark matter may distributes
on a single spatial scale with a central cusp.
In fact, such a cuspy distribution of dark matter,
called {\it universal density profile}, has been proposed by
Navarro, Frenk, \& White (1996,1997: hereafter NFW model)
based on $N$-body simulations.
The NFW model describes the dark-matter density profile as
\begin{equation}
\label{eq:NFW}
 \rho(R) = \rho_{\rm c} \delta_{\rm c} \left(\frac{R}{R_{\rm s}}\right)^{-1}
        \left(1 + \frac{R}{R_{\rm s}}\right)^{-2}~~,
\end{equation}
where 
$\rho_{\rm c}$ is the critical density that is a constant,
$\delta_{\rm c}$ is a characteristic density,
and $R_{\rm s}$ is a scale radius.
Its integral form becomes (Suto, Sasaki, \& Makino 1999)
\begin{equation}
\label{eq:NFW_MASS}
  M_{\rm DM}(R) = 4\pi\rho_{\rm c} \delta_{\rm c} R_{\rm s}^3
        \left\{ \ln\left( 1 + \frac{R}{R_{\rm s}} \right)
        - \frac{R/R_{\rm s}}{1+R/R_{\rm s}} \right\}\ .
\end{equation}
When an isothermal ICM is hydrostatically confined
in the gravitational potential corresponding to
equation (\ref{eq:NFW}) or equation (\ref{eq:NFW_MASS}),
the gas density profile is expressed as (Makino, Suto, \& Sasaki 1998)
\begin{equation}
\label{eq:NFW_ICM}
 n(R) = n_{\rm 0}
\left(1+\frac{R}{R_{\rm s}}\right)^{\frac{B R_{\rm s}}{R}}~~,
\end{equation}
where $n_{\rm 0}$ and $B$ are parameters related to
the central ICM density and the ICM temperature, respectively.
Actually, Tamura (1998) has shown
that the {\it ASCA} and {\it ROSAT} data of the Abell 1060 cluster
can be explained by a nearly isothermal ICM confined in the
NFW-type potential (possibly with an even steeper cusp).

In order to examine the interpretation in terms of the NFW model,
we again perform a simultaneous fitting to the PSPC and GIS data,
with an alternative two-phase ICM model.
In \S~6 we modeled the two emission-measure profiles,
$Q_{\rm h}(R)$ and $Q_{\rm c}(R)$, and then simulated the data.
Here we take a somewhat different approach;
while we model $Q_{\rm c}(R)$ again in terms of a
single-$\beta$ model via equation (\ref{eq:em1beta}) as before,
we specify the total mass profile instead of $Q_{\rm h}(R)$.
Specifically, we express the total mass profile
as a sum of the stellar mass profile $M_*(R)$,
and the NFW dark-matter profile $M_{\rm DM}$
given by equation (\ref{eq:NFW_MASS}).
This is because the NFW model neglects the baryonic contribution.
We fix $M_*(R)$ according to Figure~\ref{fig:masspro}a,
while we treat $\delta_{\rm c}$ and $R_{\rm s}$ in
equation (\ref{eq:NFW_MASS}) as fitting parameters.
We further assume, as in \S~6,
that the hot phase is isothermal,
the cool and hot phases are in  pressure balance,
and the cool component is distributed up to $6'$ from the center.
Then, in the outer region of $R>6'$
where the cool phase and the stellar mass are both  negligible,
the hot-phase density profile $n_{\rm h}(R)$ can be
given analytically by equation (\ref{eq:NFW_ICM}).
In contrast, within $R=6'$
where $n_{\rm h}(R)$ is affected by the cool phase and the stellar mass,
we must numerically solve the ordinary differential equation
for $n_{\rm h}(R)$ as
\begin{equation}
\label{eq:nfw_diffeq}
 \frac{{\rm d}n_{\rm h}(R)}{{\rm d}R} =
        - \frac{\mu m_{\rm p} G}{k_{\rm B} T_{\rm h}}
        \left\{ \frac{T_{\rm c}}{T_{\rm h}}
        \left(1-\frac{T_{\rm c}}{T_{\rm h}}\right)
        \frac{Q_{\rm c}(R)}{n_{\rm h}(R)}
        + n_{\rm h}(R) \right\} \frac{M_{*}(R)+M_{\rm DM}(R)}{R^2}~~,
\end{equation}
once $Q_{\rm c}(R)$, $M_{*}(R)$, and $M_{\rm DM}(R)$ are specified.
This relation refers to equation(\ref{eq:grav}),
into which equations (\ref{eq:pbal}) through (\ref{eq:meanrho}) 
have been substituted.

\begin{figure}
\hspace{1cm}(a)\hspace{7.7cm}(b)

\vspace{-10mm}
\centerline{
\psfig{file=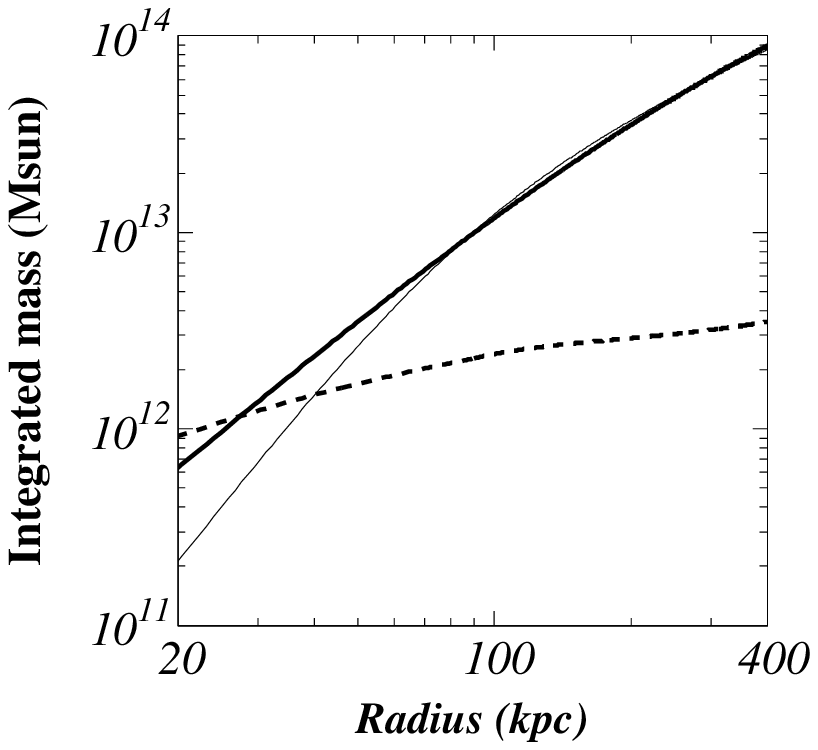,width=8.0cm,angle=0,clip=}
\psfig{file=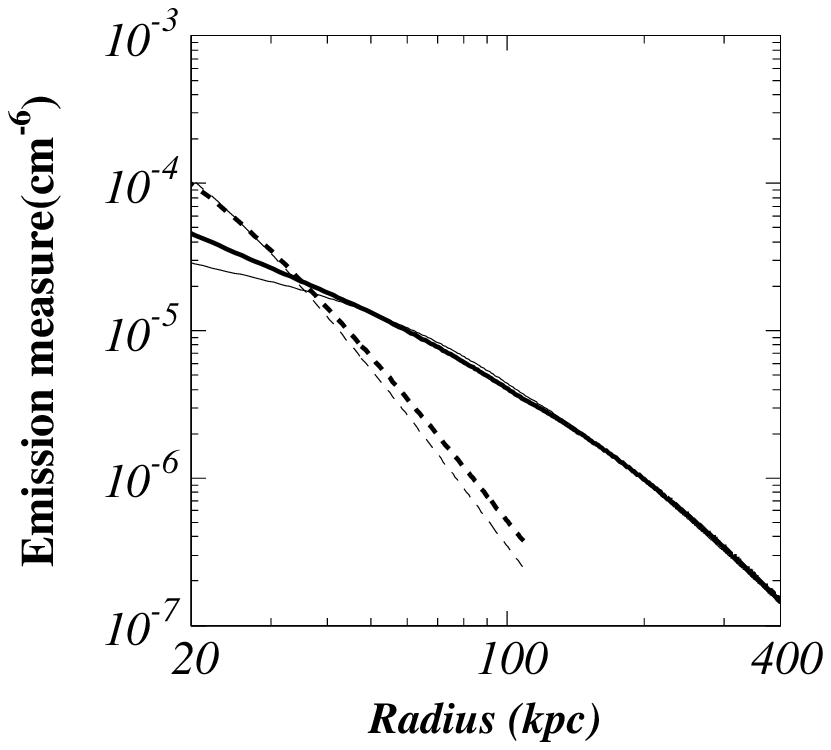,width=8.0cm,angle=0,clip=}}
\hspace{1cm}(c)

\vspace{-10mm}
\psfig{file=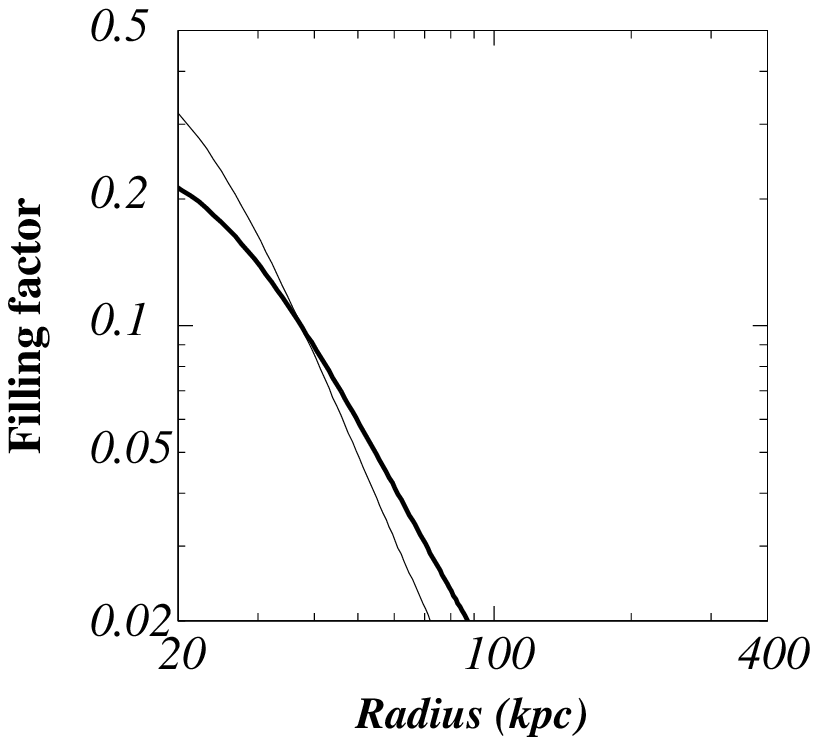,width=8.0cm,angle=0,clip=}
\figcaption{The best-fit solutions in terms of 
the Navarro, Frenk, \& White model (bold lines) 
and the King-type model (thin lines).
(a) The dark matter profiles. The stellar mass is also plotted
with the dashed line.
(b) The emission measure profiles of the hot (solid lines) 
and the cool (dashed lines) phases.
(c) The filling factor of the cool phase calculated from 
the best-fit model.
\label{fig:nfwmodel}}
\end{figure}

With this alternative ICM model,
we have fitted the GIS spectra and the PSPC profile
simultaneously as was performed in \S~6.
The hot phase temperature was left free to vary,
while the hot-phase metallicity profile was again
modeled with equation (\ref{eq:metallicity}).
The cool-phase temperature and the cool-phase metallicity
were fixed to the values derived with Model~2 in \S~5.3;
so was the overall hydrogen column density.
This has given an acceptable fit to the GIS/PSPC data,
with a reduced chi-square of $\chi^2/\nu$=400/408
(to be compared with $\chi^2/\nu=377/404$ derived
in \S~6.2 using the double-$\beta$ modeling).
The best-fit parameters are given in Table~\ref{tab:nfwking}
and the model profiles are illustrated in Figure~\ref{fig:nfwmodel}.
Although we treated both the scale radius ($R_S$) and
the characteristic density ($\delta_c$) as free parameters,
the NFW model indicates that they are correlated
in such a way that central densities are higher in lower mass systems
(Navvaro, Frenk, \& White 1996).
In order to examine the NFW-model fit result for the mass-density
relation,
we calculated the virial mass, $M_{200}$, 
which is the total mass within the sphere encompassing
a mean overdensity of 200.
The derived virial mass is
$3.7\times10^{14}$ $M_{\odot}$ for $H_0$ = 50, or 
$1.6\times10^{14}$ $M_{\odot}$ for $H_0$ = 75,
which agree with the mass-density ($M_{200}-\delta_c$) relations 
derived with
SCDM ($\Omega_0$=1, $H_0$=50) or CDM$\Lambda$ ($\Omega_0$=0.25, $H_0$=75)
cosmological models, respectively
(for detail see Navvaro, Frenk, \& White 1997).
Therefore, the dark-matter density profile of the Cen cluster
is concluded to be consistent with the NFW model.

For comparison, we also tried a King-type profile
for the dark matter distribution, instead of the NFW profile.
In this case, we substitute into equation (\ref{eq:nfw_diffeq})
\begin{equation}
M_{\rm DM}(R) = \frac{3\beta' k_{\rm B} T_{\rm h}}{\mu m_{\rm p} G}
	\frac{R^3}{R^2+R_{\rm c}'^2}~~,
\end{equation}
where the hot component temperature  $T_{\rm h}$,
the core radius $R'_{\rm c}$, and $\beta'$ are free parameters.
In the outer cluster region where the temperature structure
becomes isothermal and the stellar mass contribution is negligible,
the ICM density profile is expressed with a $\beta$-profile as
\begin{equation}
n(R) = n_{\rm 0} \left\{
        1+\left(\frac{R}{R'_{\rm c}}\right)^2
        \right\}^{-\frac{3}{2}\beta'}~~,
\end{equation}
instead of equation (\ref{eq:NFW_ICM}).
The fit with the King-type profile yielded
a minimum chi-square value of $\chi^2/\nu$=428/408 (Table~\ref{tab:nfwking}),
which is significantly worse than that with the NFW mass model.
As illustrated in Figure~\ref{fig:nfwmodel}a,
the best-fit King-type profile is less massive
than the NFW profile in the central region.

These results confirm that the dark matter mass profile
in the Cen cluster indeed deviates from a King-type one,
exhibiting a significant central excess.
In other words, the central excess in the total mass distribution,
which we have discovered, cannot be attributed entirely
to the contribution of stellar mass in the cD galaxy.
The excess dark matter can be reproduced with the single NFW profile;
then, the double-$\beta$ modeling we used in \S~4 and \S~6 may be
regarded as a conventional one rather than a physical one.
Of course, there still remains the  alternative possibility
that the dark-matter distribution in the central region in fact
has two distinct spatial scales, like in the Fornax cluster;
in that case, the double-$\beta$ modeling has a physical meaning.
In short, the central excess in the total mass profile
consists of a baryonic component and a dark component.
The latter, in turn, is either a dark halo associated with NGC~4696,
or a central cusp in the cluster-scale dark matter distribution.
With the current datasets alone,
we cannot distinguish these two alternatives.


\subsection{Nature of the central cool phase}
Our results indicate that the central volume of
the Cen cluster is permeated by the hot-phase ICM,
where a small amount of cool phase is mixed up.
The cool phase, having an average filling factor of 6--8\% within 60~kpc,
presumably takes the form of blobs or filaments immersed in the hot phase
on angular scales too small to be resolved with {\it ASCA}.
Indeed, the {\it ROSAT} HRI observations of some cD clusters have revealed
evidence of small-scale (a few kiloparsec) X-ray inhomogeneities
(Sarazin, O'Connell, \& McNamara 1992ab; Prestwitch et al. 1995).
Then, what is the nature of the cool phase?
The failure of the cooling-flow model (\S~5.4) inspires
us to explore an alternative interpretation.

First of all, the cool phase has a temperature of $T_{\rm c} = 1.4$ keV,
and a core radius of $5-27$ kpc centered on the cD galaxy.
These values are similar to, or slightly higher (larger) than,
those of inter-stellar medium (ISM) of ordinary
elliptical galaxies (e.g. Forman, Jones, \& Tucker 1985;
Awaki et al. 1994; Matsumoto et al. 1997).
It is hence suggested that the cool phase has a close connection
to the ISM of the cD galaxy (FEA94; Makishima 1994, 1996b, 1997).
Furthermore, the value of $T_{\rm c}$ is comparable
to the virial temperature of the central dimple
in the gravitational potential.

Employing the two-phase picture
together with equations (\ref{eq:pbal}) and (\ref{eq:emdens}),
and assuming the cool-phase abundance of 1.0 solar,
the total emission integral and total mass of the cool phase
are calculated to be $6\times10^{65}$ cm$^{-3}$
and $(4-5) \times 10^{10} \; M_{\odot}$, respectively.
(Note that the cool-phase mass derived in FEA94,
$(1-2) \times 10^{11} \; M_{\odot}$, was subject to a trivial mistake.)
These values are comparable to those of the ISM associated with
the most luminous non-cD elliptical galaxies
(Forman et al. 1985; Matsushita 1997).
As a consequence, the cool phase of the Cen cluster lies on a smooth
extension of the temperature vs. emission-integral relation
or temperature vs. gas-mass relation seen among non-cD
elliptical galaxies (Matsumoto et al. 1997; Matsushita 1997).
The ratio of the cool-phase mass
to the stellar mass of NGC~4696 is $2.0-2.5$\%,
which is also comparable to those of non-cD elliptical
galaxies (Forman et al. 1985).
Therefore, the cool phase may be regarded
essentially as the ISM of the cD galaxy.

In comparison with the ISM of the most X-ray luminous non-cD ellipticals,
which is thought to have a filling factor of $f \sim 1$,
the cool phase of the Cen cluster may have a similar gas mass
and a similar spatial extent with $f \sim 0.07$ on average.
Then, we expect the cool phase of the Cen cluster to be
$0.07^{-1} \sim 14$ times more luminous
than the ISM of non-cD ellipticals.
Actually, the 0.5--4 keV luminosity of the cool phase
of the Cen cluster, $1.1 \times 10^{43}$ erg s$^{-1}$,
is some 10 times higher than those of the most X-ray luminous
ordinary elliptical galaxies (e.g. Forman et al. 1985),
in good agreement with the above prediction.

In \S~5.3, we have determined the iron abundance
of the cool phase to be $\sim 1.0$ solar.
This should be compared with ISM metallicity
of non-cD elliptical galaxies.
Through observations with {\it ROSAT} (Forman et al. 1993),
{\it BBXRT} (Serlemitsos et al. 1993), and {\it ASCA}
(Awaki et al. 1994; Loewenstein et al. 1994; Matsumoto et al. 1997),
the ISM of elliptical galaxies were found to be extremely metal poor.
This was a serious theoretical problem,
because the measured ISM abundances, often below half a solar,
fall even below those
implied by the contribution from stellar mass loss,
leaving essentially no room for the supernova contribution
(Arimoto et al. 1997).
However, the detailed recent work by Matsushita (1997)
and Matsushita et al. (1997) clearly indicates
that the most X-ray luminous ellipticals
have the ISM abundances of 0.7--1.0 solar,
in agreement with those from the stellar mass-loss.
Then, the cool-phase metal-abundance of the Cen cluster
that we measured is close to, or only slightly higher than,
those of the most X-ray luminous non-cD ellipticals.

Thus, from every aspect including the temperature,
angular extent, mass, and metallicity,
the cool phase of the Cen cluster can be regarded
as the ISM associated with the cD galaxy, NGC~4696,
which is among the most luminous ellipticals.
The only basic difference is
that the ISM of NGC~4696 is confined 
not only by the central potential drop
but also by external pressure from the surrounding hot phase 
to achieve the considerably low filling factor,
hence the higher density and the higher luminosity,
than those of ordinary ellipticals.

\subsection{Origin of the metallicity increase}

Although the increased metallicity in the cool phase has been explained
successfully in \S~7.3 using analogy to the ISM of ordinary ellipticals,
we must also explain the iron concentration seen in the hot phase.
Indeed, the metallicity increase in the central part of the hot phase
is even more robust than the abundance enrichment in the cool phase,
because the former can be directly evidenced
by the central increase in the Fe-K line equivalent width.
Such an effect is not limited to the Cen cluster,
because cD clusters generally exhibit statistically significant
metallicity increases at the center (Fukazawa 1997).

\begin{figure}
\centerline{
\psfig{file=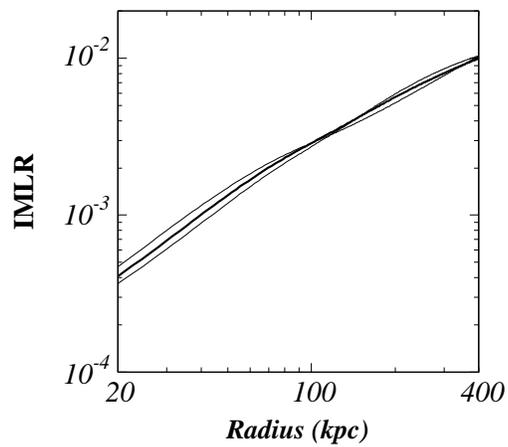,width=8.0cm,angle=0,clip=}}
\figcaption{The Iron Mass to Light Ratio derived with the 2-phase model
fit.
The thick line shows the best-fit model,
while thin lines show extreme cases within 90\% ($\Delta \chi^2$=2.706)
confidence limits.
\label{fig:imlr}}
\end{figure}

Obviously, the metals in the ICM must have been supplied
by individual stars in the member galaxies,
as clearly demonstrated by the large-scale abundance gradient
found by Ezawa et al. (1997) from the AWM7 cluster.
Therefore, a comparison between the spatial distribution
of the iron mass and the optical light profile is crucial.
In Figure~\ref{fig:imlr}, 
we plot iron-mass to light ratio (IMLR; Ciotti et al. 1991),
i.e. ratio of the total iron mass (contained in both ICM phases) within
each radius to the total optical luminosity within the same radius,
expressed in solar units.

The plot clearly reveals an outward increase in the IMLR,
until it reaches $\sim 0.01$ (at $\sim 400$ kpc)
which is a typical IMLR found in clusters of galaxies
(Ciotti et al. 1991; Arnaud et al. 1991; Tsuru 1992;
Renzini et al. 1993; Renzini 1997).
In other words, the iron is more widely spread than the stars
contrary to the impression we get from the measured steep 
metal profile,
while the iron is more concentrated than the ICM mass profile
as evidenced by the strong central metallicity enhancement.
Actually, another two-phase model, in which the constant IMLR is assumed
rather than equation~(\ref{eq:metallicity}) is adopted,
gave a significantly worse fit to the GIS+PSPC data with
$\chi^2/\nu$=464/407,
because it predicts the iron concentration much stronger than is observed.
Note that this argument utilizes only the observed quantities,
and unaffected by uncertainties in the supernova rate or its time
evolution.

Reisenegger et al. (1996) argued that the central metallicity increase,
produced by the cD galaxy, may be reduced by the cooling flow,
because the metal-rich plasma is swept inwards and then disappears.
However, the value of $R_{\rm Z} \sim 4'$ (Table~\ref{tab:ringspecfit})
we have obtained is significantly larger than the scale length of 
the metallicity increase, $\sim 2'$, calculated by
Reisenegger et al. (1996).
(Note that the instrumental effects have already been removed from 
$R_{\rm Z}$.)
Furthermore, the prediction by Reisenegger et al. (1996),
that the CF rate should negatively correlate with the
metal concentration scaled to the stellar mass of the cD galaxy,
has not been confirmed in a sample of poor clusters
observed with {\it ASCA} (Tamura et al. 1996).
Therefore, the outward increase in IMLR
need to be explained by some mechanism other than the CF:
there are the following two possible mechanisms.
 
First, the supernova ejecta may be so energetic
that they may not be confined by the gravitational potential of NGC~4696.
As a supporting evidence, we refer to a large number of
IMLR measurements in clusters, groups, and elliptical galaxies
(Ciotti et al. 1991; Arnaud et al. 1991; Tsuru 1992;
Renzini et al. 1993; Renzini 1997; Matsushita 1997).
For clusters having the X-ray temperature higher than $\sim$1~keV,
the IMLR integrated over the whole system is remarkably constant at 
$\sim0.02$.
On the contrary, the IMLR of systems that have temperature less than
$\sim$1~keV is significantly lower on average, and scatters widely.
As the above authors argue, these results indicate
that the metal-rich gas escapes significantly from smaller systems,
such as groups and single elliptical galaxies,
because their gravitational potential is not deep enough.
Then, a considerable fraction of the supernova products in NGC~4696
may well have escaped from its gravitational potential in the form of
winds, to spread over the cluster space.
 
In addition, motion of the cD galaxy can also contribute to the metal
spread-out.
Although NGC~4696 now seems to sit at the bottom of the cluster potential,
it may have been moving through the cluster space
and gradually became stationary over its lifetime.
Then, the metal-enriched supernova products may have been
continually spilling over the galaxy potential,
or stripped off the host galaxy by ram pressure exerted from the ICM,
to be left over in the surrounding ICM as the galaxy moves away.
Actually, the position of a cD galaxy,
which generally coincides with the X-ray peak position,
is often displaced from the centroid of the large scale X-ray emission
(e.g. Ikebe et al. 1996;  Neumann \& B\"{o}hringer 1995;
Lazzati \& Chincarini 1998).
Offsets are seen also between the radial velocities of the central galaxies
and the average velocities of the member galaxies
(e.g. Beers \& Geller 1983;
Zabludoff, Huchra, \& Geller 1990; Zabludoff et al. 1993).
This suggests that the central galaxy is in an
harmonic oscillation mode within the central potential minimum
(Lazzati \& Chincarini 1998),
or that the central potential structure itself is still evolving with time.
 
 From these considerations, we can interpret that
the excess iron in the central region of the Cen cluster
was actually produced by the cD galaxy over the Hubble time,
and transported to a wider cluster space.
 

\subsection{Radiative cooling}
 
We have so far argued that the cool phase is
essentially the ISM associated with the cD galaxy.
However, our two-phase solution implies the central plasma densities
of $\sim2 \times 10^{-2}$ and $\sim6 \times 10^{-2}$ cm$^{-3}$,
hence radiative cooling times of $\sim1$ Gyr and $\sim8\times10^7$ yr,
for the hot and cool phases respectively (both assuming 1 solar abundance).
Thus, the cool phase would be subject to a strong radiative cooling;
the stable presence of such a plasma component must be explained.
 
Before attempting to solve this problem,
let us consider ISM of non-cD elliptical galaxies.
In many elliptical galaxies,
the ISM has been observed to be hotter than the stellar kinetic temperature
by up to a factor of $\sim 2$ (Matsumoto et al. 1997; Matsushita 1997).
Therefore, the presence of some ubiquitous ISM heating mechanism is
indicated.
This tendency becomes more prominent
in X-ray fainter ellipticals (Matsumoto et al. 1997),
presumably because the heat input is less diluted
and the ISM cooling effect becomes less important
in the X-ray fainter galaxies, which have smaller amount of ISM.
At the same time, the measured ISM metal abundances 
clearly decrease towards X-ray fainter ellipticals 
(Matsumoto et al. 1997; Matsushita 1997).
Therefore, the supernova heating cannot be the main mechanism 
of the ISM heating.

We presume that the ISM is heated by random stallar motion in each galaxy,
through amplification of interstellar magnetic fields.
The fields will in turn reconnect to heat up the ISM,
in particular near the magnetopause of each star.
Quantitatively, the random stellar motion in giant elliptical galaxies
has a total kinetic energy of $(0.1-1) \times 10^{60}$ ergs.
If a few tens percent of this energy is spent in the ISM heating over
the Hubble time,
we can expect a heating luminosity up to 
a few times $10^{41}$ ergs s$^{-1}$,
which is comparable to the observed X-ray luminosities of elliptical
galaxies.
As argued by Kritsuk (1992, 1996, 1997),
this heating mechanism may balance the radiative cooling of the ISM,
and keep the ISM in an apparent steady state.
(Microscopically, the ISM may recycle between the cooling and heating
stages.)
This scenario is best supported by the observational fact
that the ISM emissivity is spatially proportional to the
stellar light density (Forman, Jones, \& Tucker 1985),
the latter being proportional to the local heat input.
 
If we apply the same idea to the cool phase of the Cen cluster,
its high luminosity ($1 \times 10^{43}$ ergs s$^{-1}$) requires
that nearly all the stellar kinetic energy of NGC~4696
has been released in the plasma heating.
In other words, we must invoke a significant dynamical evolution of
NGC~4696.
This is actually quite plausible, because NGC~4696,
being a cD galaxy, must have grown up via cannibalism.
Conversely, the merger process would not proceed without such energy 
dissipation,
and the plasma heating is one of the most plausible dissipation mechanisms.
Suppose that the stellar mass of NGC~4696 has doubled to the present value
by repeated mergers of dwarf galaxies over the Hubble time.
The total energy release will be $\sim 1 \times 10^{42}$ ergs s$^{-1}$
if the infalling dwarf galaxies are assumed to have no dark-matter halos.
When the dwarfs are thought to have mass-to-luminosity ratios
similar to that of the present-day NGC~4696,
the energy release will be $\sim 5$ times larger.
Then, we expect an average heating luminosity in the range
$(1-5) \times 10^{42}$ ergs s$^{-1}$,
which may be sufficient to sustain the cool-phase luminosity.

Finally, we must consider the spatial configuration of the cool and hot
phases.
If they were not thermally insulated from each other,
the cool phase would be heated up and disappear quickly
by conductive heat flux from the hot phase (Takahara, \& Takahara 1979).
Therefore, the two phases must be thermally insulated from each other,
presumably by magnetic fields.
For example, the cool phase may fill up the interior of
numerous magnetic loops anchored to the cD galaxy (Makishima 1997),
whereas the hot phase may permeate the open-field regions
which connect to the outer cluster space.
The heating process considered above in principle operates on both phases,
but it must be much more effective on the cool phase
due to the higher field strengths inside each magnetic loop
and due to the closed magnetic configuration.
Any heating effect on the hot phase would be difficult to observe,
because the heat deposit would be quickly transported to the
outer region by the heat conduction within the hot phase,
with a typical time scale of $\sim 10^7$ yr.

In summary, by invoking the merger process,
magnetic reconnection, and a certain magnetic configuration,
we can at least qualitatively give a self-consistent account of the
two-phase nature observed in the central region of the Cen cluster.
 
\acknowledgments
We are grateful to Professor J. Tr\"{u}mper and the {\rm ROSAT} team
for support and providing us with the {\it ROSAT} All Sky Survey data.
We also thank H. B\"{o}hringer and T. Reiprich for discussion and
helping the {\it ROSAT} data analysis.
We appreciate the {\it ASCA} team for operating the spacecraft
and supporting the data analysis.
Y.I. is supported by the post-doctoral program of Max-Planck Gesellschaft.
K.Matsushita is supported by the post-doctoral program of Japan Society
for the Promotion of Science.

 
\begin{table}
\begin{center}
\caption{Log of {\it ASCA} observations of the Centaurus cluster.}
\label{tab:obslog}
\begin{tabular}{llcccc}
\hline\hline
 & date & \multicolumn{2}{c}{pointing position (J2000)} \
& exposure time\ $^{a)}$ \\
 &      &   R.A.(deg)   &   DEC.(deg)   & (ksec)         \\
\hline
1 & 1993 June 30 & 192.47 & --41.23 & 16.7  \\
2 & 1993 July 5  & 192.55 & --40.96 & 13.2  \\
3 & 1993 July 5  & 192.79 & --41.32 & 8.2  \\
4 & 1995 July 19 & 192.09 & --41.32 & 58.6  \\
\hline
\end{tabular}
\end{center}
\begin{itemize} 
\setlength{\itemsep}{-2mm}\setlength{\itemindent}{-5mm}
  \item[$^{a)}$]: The total exposure time surviving the data screening
	process.
\end{itemize}
\end{table}

\begin{table}
\begin{center}
\caption{The results of fitting radial count-rate profiles with single and
double $\beta$ models.} 
\label{tab:rprofit}
\begin{tabular}{cl|ccc|cccc}
\hline\hline
Detector  & Energy &  \multicolumn{3}{c}{Single $\beta$ model} \
& \multicolumn{4}{|c}{Double $\beta$ model\ $^{ a)}$} \\
     & (keV)    & $R_{\rm c}(')$ & $\beta$ & $\chi^2/\nu$ \
     & $R_{\rm c,2}(')\ ^{ b)}$  & $\beta_{\rm 2}\ ^{ b)}$ \
     & $\epsilon_{0,2}/\epsilon_{0,1}\ ^{ c)}$  & $\chi^2/\nu$ \\
\hline
PSPC & 0.2--2.0 & 1.2 & 0.47 & 139/116 \
  & 1.5$^{+4.5}_{-1.3}$ & 0.92$^{+3.8}_{-0.45}$ & 124 & 96.0/115 \\
($r>6'$) & 0.2--2.0 & 7.3$^{+2.2}_{-2.0}$ & 0.57$\pm0.04$ & 91.8/110 \
  &  &  &  & \\
GIS  & 1.6--2.4 & 1.0 & 0.43 & 92.7/21 \
  & 2.0$^{+1.6}_{-0.8}$ & 0.98$^{+0.83}_{-0.29}$ & 52.8 & 35.0/20 \\
     & 2.4--4.5 & 1.5 & 0.44 & 89.9/21 \
  & 2.0$^{+1.7}_{-1.0}$ & 0.86$^{+0.74}_{-0.27}$ & 31.6 & 25.0/20 \\
     & 4.5--6.0 & 2.4 & 0.45 & 46.7/21 \
  & 1.9$^{+3.5}_{-0.8}$ & 0.69$^{+1.42}_{-0.16}$ & 17.4 & 16.4/20 \\
     & 6.0--7.1 & 1.9 & 0.47 & 41.3/21 \
  & 1.9$^{+1.6}_{-0.7}$ & 0.63$^{+0.46}_{-0.12}$ & 30.4 & 28.4/20 \\
     & 7.1--10.0& 1.9 & 0.43 & 17.1/21 \
  & 4.0$^{+\infty}_{-3.2}$ & 1.54$^{+\infty}_{-1.04}$ & 10.5 & 13.5/20 \\
\hline
\end{tabular}
\begin{itemize} 
\setlength{\itemsep}{-2mm}\setlength{\itemindent}{-5mm}
  \item[$^{a)}$]: The core radius and $\beta$ parameter of 
	the wider $\beta$-component are fixed to $7'.3$ and 0.57,
	respectively, which are obtained by the single $\beta$-model fit
	to the PSPC profile outside 6 arcmin.
  \item[$^{b)}$]: Shape parameters of the narrower $\beta$-component of
	equation (2).
  \item[$^{c)}$]: The normalization ratio between the wider and narrower 
	$\beta$-component of equation (2).
\end{itemize}
\end{center}
\end{table}

\begin{table}
\begin{center}
\caption{The spectral fitting within the central $3'$.} 
\label{tab:specfit}
\begin{tabular}{lccccccc}
\hline\hline
Model  & $\chi^2/\nu$ & $T_{\rm c}$ & $T_{\rm h}$ & $EI_{\rm h}\,^{b)}$
& $N_{\rm H}$ & $\Delta N_{\rm H}\,^{c)}$ & $EI_{\rm c}\,^{d)}$ \\
&  & (keV) & (keV) &  & \multicolumn{2}{c}{($10^{21}$ cm$^{-2}$)} &or $\dot{M}\,^{e)}$ \\
\hline
1        & 133/123 & 1.4$\pm0.1$            & 4.4$^{+0.3}_{-0.2}$ \
               & 9.2     & 0.98$^{+0.14}_{-0.13}$ & \ & 8.3 \\
2$^{a)}$ & 119/121 & 1.4$^{+0.3}_{-0.1}$    & 4.4$^{+0.6}_{-0.4}$\
               & 9.1     & 0.97$^{+0.14}_{-0.13}$ & \ & 7.5\\
3$^{a)}$ & 118/120 & 1.4$^{+0.3}_{-0.2}$    & 4.2$^{+0.8}_{-0.3}$\
               & 9.8     & 0.74$^{+0.30}_{-0.11}$ & 1.1$^{+1.5}_{-1.1}$ \ & 7.0\\
4$^{a)}$ & 159/122 & & 4.7 & 0.59  & 1.3 & & 38 \\
5$^{a)}$ & 142/121 & & 4.3 & 0.31  & 0.0 & 1.9 & 38 \\
\hline
\end{tabular}
\end{center}
\begin{itemize} 
\setlength{\itemsep}{-2mm}\setlength{\itemindent}{-5mm}
  \item[$^{a)}$]: The parameters of the additional emission line at
	$\sim1.7$~keV are quite similar among these different spectral models.
	The central energy and equivalent width of the line are
   	in the range of 1.72--1.74~keV and 17--22~eV, respectively.
  \item[$^{b)}$]: The emission integral of the hotter component, 
	in units of $10^{65}$ cm$^{-3}$.
  \item[$^{c)}$]: The additional absorption for the cooler component
	or the cooling flow component.
  \item[$^{d)}$]: The emission integral of the cooler component 
	in $10^{65}$ cm$^{-3}$, for models 1 through 3.
  \item[$^{e)}$]: The mass deposition rate of the cooling flow 
	in $M_\odot$ yr$^{-1}$, applicable to models 4 and 5.
\end{itemize}
\end{table}

\begin{table}
\hspace{-2cm}
\begin{flushleft}
\caption
{Results of the simultaneous fits to the 10 GIS ring-sorted spectra
and the PSPC radial profile.}
\label{tab:ringspecfit}
\begin{tabular}{cccccccc}
\hline\hline
$T_{\rm h}$ & $R_{\rm c,1}$ & $\beta_1$ & $R_{\rm c,2}$ & $\beta_2$ \
& $\Delta Z$ & $R_{\rm Z}$ & $Z_0$   \\
  (keV)      & (arcmin)    &           &             &             \
& (solar)    & (arcmin)    & (solar) \\
\hline
3.94 & 7.5 & 0.57 & 7.5 & 3.37 & 1.5 & 4.1 & 0.20 \\
$\pm0.07$ & $^{+2.1}_{-0.8}$ & $^{+0.04}_{-0.02}$ \
& $^{+\infty}_{-5.2}$ & $^{+\infty}_{-2.77}$ & $^{+1.0}_{-0.4}$ \
& $^{+3.1}_{-1.7}$ & $^{+0.08}_{-0.11}$ \\
\hline
\end{tabular}

\begin{tabular}{ccccc}
\hline\hline
$T_{\rm c}$ & $Z_{\rm c}$ & $R_{\rm c,c}$ & $\beta_{\rm c}$ & $\chi^2/\nu$
\\
(solar)     &    (keV)    &  (arcmin)     &                 &             
\\
\hline
1.4  &  1.0  &  0.5       & 0.57       & 377/405 \\
fix  &  fix  & $^{+0.5}_{-0.3}$ & $^{+0.17}_{-0.11}$ &         \\
\hline
\end{tabular}
\begin{itemize} 
\setlength{\itemsep}{-2mm}\setlength{\itemindent}{-5mm}
  \item[]: The ratios among the normalizations of equation (4) and (6) are 
	$n_{0,1}:n_{0,2}:n_{0,c}=1:2.26:14.1$.
\end{itemize}
\end{flushleft}
\end{table}

------------------------------------------------------
\begin{table}
\begin{center}
\caption{Summary of the two-phase model}
\label{tab:summary}
\begin{tabular}{ccc}
\hline\hline
                   & hot component               & cool component   \\
\hline
temperature (keV)  &  3.9                        &   1.4  \\
$\beta$            &  0.59                       &   0.57 \\
core radius (arcmin) & 9.6/2.4                   &   0.6  \\
central density (cm$^{-3}$) & $\sim0.02$         &   $\sim0.06$       \\
metallicity 1      & $\Delta Z = 1.5$            &   $Z_{\rm Fe}=1.0$ \\
metallicity 2      & $R_{\rm Z} = 4'.1$          &  \\
metallicity 3      & $Z_0 = 0.2$                 &  \\
\hline
\end{tabular}
\end{center}
\end{table}

\begin{table}
\begin{center}
\caption{The Results with the alternative two-phased ICM models.}
\label{tab:nfwking}
\begin{flushleft}
\begin{tabular}{cccc}
\hline\hline
 & $R_{\rm s}$/$R'_{\rm c}$   & $\delta_c$ $^{c)}$  & $\beta'$ \\
 & (arcmin)                   &                     &          \\
\hline
NFW $^{a)}$  & 17.4 & $1.25\times10^{4}$ & \\
KING $^{b)}$ & 4.9  & & 0.51 \\
\hline
\end{tabular}
\begin{tabular}{ccccccc}
\hline\hline
$T_{\rm h}$ & $\Delta Z$ & $R_{\rm z}$ & $Z_0$ & $R_{\rm c,c}$ & 
$\beta_{\rm c}$ & $\chi^2/\nu$ \\
(keV)       & (solar)    & (arcmin)    & (solar) & (arcmin) & & \\
\hline
3.97 & 1.54 & 4.0 & 0.20 & 1.0 & 0.66 & 400/408 \\
3.95 & 3.83 & 1.8 & 0.28 & 0.9 & 0.70 & 428/408 \\
\hline
\end{tabular}
\end{flushleft}
\end{center}
\begin{itemize} 
\setlength{\itemsep}{-2mm}\setlength{\itemindent}{-5mm}
  \item[$^{a)}$]: The dark matter distribution is modeled with the
	Navarro, Frenk, \& White model.
  \item[$^{b)}$]: The King-type dark matter model.
  \item[$^{c)}$]: In another expression, 
	$\rho_c \delta_c$ = $0.52\times10^{10}$ ($M_{\odot}$arcmin$^{-3}$).
\end{itemize} 
\end{table}

\clearpage

\end{document}